\documentclass{article}
\usepackage[utf8]{inputenc}
\usepackage[UKenglish]{babel}
\usepackage[english]{varioref}
\usepackage[nottoc]{tocbibind}
\usepackage[toc,page]{appendix}
\usepackage{caption}
\usepackage{amsmath}
\usepackage{amsthm}
\usepackage{amssymb}

\usepackage{mathrsfs}
\usepackage{geometry}
\usepackage{float}
\usepackage[caption = false]{subfig}

\usepackage{graphicx,float}
\usepackage{array}

\usepackage{hyperref}
\hypersetup{
    colorlinks = True,
    linkcolor = blue,
    citecolor = blue
}
\usepackage[font = small,labelfont = bf]{caption}
\usepackage{booktabs}

\usepackage{csquotes}

\usepackage{authblk}
\usepackage{graphbox}

\usepackage{titlesec}
\titleformat*{\section}{\itshape}
\titleformat*{\subsection}{\itshape}
\titleformat*{\subsubsection}{\itshape}

\usepackage{float}
\usepackage[caption = false]{subfig}
\usepackage{subfiles}
\usepackage{textcomp}
\usepackage{mathcomp}
\usepackage[table]{xcolor} % http://ctan.org/pkg/xcolor
\usepackage{pythonhighlight}

\begin{document}

\title{Infra-red thermographic inversion in ST40}

\author[1,*]{M. Moscheni}
\author[1]{E. Maartensson}
\author[1]{M. Robinson}
\author[1]{C. Marsden}
\author[1]{A. Rengle}
\author[3]{A. Scarabosio}
\author[1]{P. Bunting}
\author[2]{T.K. Gray}
\author[1,4]{S. Janhunen}
\author[1]{E. Vekshina}
\author[1]{X. Zhang}
\author[1]{the ST40 Team}

\affil[1]{Tokamak Energy Ltd., 173 Brook Drive, Milton Park, Abingdon, United Kingdom}
\affil[2]{Oak Ridge National Laboratory, Oak Ridge, TN USA}
\affil[3]{LINKS Foundation, Via Pier Carlo Boggio 61, Torino, 10138, Italy}
\affil[4]{Current address: Los Alamos National Laboratory, New Mexico, USA}
\affil[*]{Corresponding author: mosca\_matteo@hotmail.com}

\date{}
\setcounter{Maxaffil}{0}
\renewcommand\Affilfont{\itshape\small}

\maketitle

%%%%%%%%%%%%%%%%%%%%%%%%%%%%%%%%%%%%%%%%%%%%%%%%%%%%%%%%%%
%%%%%%%%%%%%%%%%%%%%%%%%%%%%%%%%%%%%%%%%%%%%%%%%%%%%%%%%%%

\section*{Abstract}

Infra-red (IR) thermography is an essential diagnostic tool for understanding the edge plasma behavior in fusion devices. In this work, we present a new in-house numerical tool, Functional Analysis of Heat Flux (FAHF), for IR thermographic inversion on Tokamak Energy's spherical tokamak (ST40). FAHF, written in Python, is designed for multi-2D thermographic inversions by solving the heat conduction equation within the divertor tiles using the finite difference method, and an explicit time stepping scheme. Utilising IR camera data with the highest available effective spatial resolution, FAHF calculates the plasma perpendicular heat flux density on the divertor tile surfaces $-$ a crucial quantity for edge plasma analysis. The tool's internal numerics is first verified through formal time and space convergence analyses, and further corroborated by an energy balance assessment. Although FAHF demonstrates significant sensitivity to user-selected spatial resolution, precise heat flux values are recoverable by ensuring a sufficiently high resolution. Implications for the optimal resolution of both the code and the diagnostic system are discussed. Finally, FAHF's model and geometry simplifications are confirmed to be accurate within 10\%, based on comparison with COMSOL Multiphysics\textsuperscript{\textregistered} simulations. As such, FAHF is proven to be a precise and accurate tool for IR thermographic inversions in ST40.
\\\\{\normalfont\textit{Keywords}: ST40, infra-red thermography, thermographic inversion, heat flux, scrape-off layer plasma, finite difference method.}

%%%%%%%%%%%%%%%%%%%%%%%%%%%%%%%%%%%%%%%%%%%%%%%%%%%%%%%%%%
%%%%%%%%%%%%%%%%%%%%%%%%%%%%%%%%%%%%%%%%%%%%%%%%%%%%%%%%%%

\section{Introduction}\label{sec:introduction}

Tokamak-based nuclear fusion presents numerous physico-technological challenges. Among them, a robust power exhaust solution which maintains the integrity of plasma-facing components (PFCs) is critical\cite{donne2019}. Understanding the scrape-off layer (SOL) transport \cite{stangeby2000plasma} is essential for establishing such a robust solution. In pilot plants \cite{step, fnsf}, unmitigated heat flux densities of hundreds of $\text{MW} \times \text{m}^{-2}$ are expected. Despite the mitigation strategies routinely undertaken in tokamaks \cite{subbaNF2021, Kallenbach_2013, PITTS2019100696}, the current PFC technologies\cite{PITTS201760} are only effective up to tens of $\text{MW} \times \text{m}^{-2}$, therefore necessitating a reliable determination of heat flux parameters such as peak heat flux $q_0$ and power fall-off length $\lambda_q$ for appropriate designs.
\\Infra-red (IR) thermography\cite{ir} emerges as a prime diagnostic tool for characterising the SOL plasma and determining these heat flux parameters. The evolution in time of the PFC surface temperature, measured via the IR thermography, informs the IR thermographic inversion, which involves the solution of the heat conduction equation\cite{incropera}. Although this is a well-established method, complexities such as space and temperature dependencies of PFC material properties, and their intrinsic three-dimensionality\cite{heat} complicate modelling efforts, also by lengthening the required computational time.
\\Numerous teams have significantly contributed to the development of IR thermographic systems and inversion techniques. Among them are: Alcator C-Mod\cite{CMOD_IR}, ASDEX\cite{ASDEX_IR_1, ASDEX_IR_2, ASDEX_IR_3}, COMPASS\cite{COMPASS_IR}, DIII-D\cite{DIIID_IR, Adebayo-Ige_2024}, EAST\cite{EAST_IR}, ITER\cite{ITER_IR_1}, JET\cite{JET_IR_1, JET_IR_2}, KSTAR\cite{KSTAR_IR_1, KSTAR_IR_2}, LHD\cite{LHD_IR}, MAST\cite{MAST_IR_1, MAST_IR_2}, NSTX\cite{NSTX_IR, Adebayo-Ige_2024}, SPARC\cite{SPARC_IR_1, SPARC_IR_2}, TCV\cite{TCV_IR_1, TCV_IR_2}, W7-X\cite{W7X_IR}, and WEST\cite{WEST_IR_1, WEST_IR_2, WEST_IR_3}. 
\\Recent experiments, including those on the ST40 spherical tokamak\cite{zhang2024experimental}, owned and operated by Tokamak Energy Ltd., COMPASS\cite{COMPASS_lq} and TCV\cite{TCV_lq} suggest that predictions from existing scalings\cite{Eich_2013} may overestimate $\lambda_q$ in certain scenarios, hence rising concern on the appropriate spatial resolution of the diagnostic system. An analysis of the above-mentioned references indicates that the hardware specifications at times only marginally meet, or fail to meet, the resolution requirements needed to accurately diagnose $\lambda_q$, even for values predicted by the existing scalings\cite{Eich_2013}.
\\In this work, we introduce the high-resolution IR thermography system of ST40 and FAHF (Functional Analysis of Heat Flux), a new in-house Python-based flexible tool for high-resolution IR thermographic inversions on ST40. FAHF addresses the afore-mentioned modelling challenges with a fast, precise and accurate approach for solving the heat conduction equation. It is developed with prime emphasis on speed and modularity, which allows model extensions and applications to different cases.
\\FAHF's precision is verified through formal convergence analyses. The accuracy of the tool is then estimated via COMSOL Multiphysics\textsuperscript{\textregistered}, by means of a cross-code comparison, a proven method for quantifying model and geometry errors \cite{Moscheni_2022, Moscheni_2024}, paramount in ST40. We also demonstrate that ST40's IR thermography is particularly well-suited for diagnosing both a narrow\cite{zhang2024experimental} $\lambda_q$ and fine-scale structures in the heat flux footprint. 
\\The paper is organised as follows. Sec. \ref{sec:experimental} expounds on the ST40 tokamak, experimental setup and data selection. The fundamentals of the thermographic inversion are then described in Sec. \ref{sec:theory_general}. The numerical methods and tools adopted in this work are introduced in Sec. \ref{sec:methods}. The results of the study are collated in Sec. \ref{sec:results}, and discussed in Sec. \ref{sec:discussion}. Lastly, Sec. \ref{sec:conclusions} summarises the conclusion of the work, and comments on future studies.

%%%%%%%%%%%%%%%%%%%%%%%%%%%%%%%%%%%%%%%%%%%%%%%%%%%%%%%%%%
%%%%%%%%%%%%%%%%%%%%%%%%%%%%%%%%%%%%%%%%%%%%%%%%%%%%%%%%%%

\section{Experimental setup}\label{sec:experimental}

    This section aims at detailing the relevant features of the ST40 tokamak, of its divertor and of the IR thermographic apparatus. With the identification of a suitable plasma shot in the ST40 database, this ultimately constitute a solid foundation for a high-resolution thermographic inversion. 

    \subsection{The ST40 tokamak}\label{sec:st40}
    
        ST40 \cite{McNamara_2024, McNamara_2023} is a compact, high-field spherical tokamak, with a major radius $0.4 - 0.5$ m, minor radius $0.2 - 0.25$ m, and a magnetic field up to 2.0 T on axis.
        \\Disconnected double-null diverted plasmas \cite{Brunner_2018, Osawa_2023} are most commonly featured in ST40. Fig. \ref{fig:bundle-a} shows the EFIT magnetic reconstruction \cite{efit} in the poloidal cross section for shot 11376 at time instant 128 ms (Sec. \ref{sec:ref_shot}). Its primary (solid) and secondary (dashed) separatrices are distinct, and displayed in magenta. In Fig. \ref{fig:bundle-a}, the secondary separatrix only contacts the wall (black) with its divertor legs above the X-point, leaving enough clearance to the centre column ($R = 0.17$ m).
        \\ST40 is provided with highly-shaped, non-axisymmetric, fish-scaled\cite{heat} tiled divertor targets\cite{marsden2024}, pictured in Fig. \ref{fig:bundle-b}, with inter-tile gaps of 4.5 mm in width, and no active cooling. The leading edge of each tile is poloidally chamfered\cite{heat}, an area excluded in the present study. The leftover volume of the divertor tiles is thus a rectangular cuboid, with 4 mm of plasma-facing TZM molybdenum (Mo) brazed onto 25 mm of copper-cromium-zirconium (CuCrZr).
        \\Poloidally, the geometry of the divertor is that of a short-legged, open, non-baffled divertor, with the inner leg orthogonal to the plate, and an horizontal outer leg (Fig. \ref{fig:bundle-a}). This configuration suppresses the role played by neutral particles at the outer target\cite{cowley2024, verhaegh2024, Sun_2023}, hence favouring attached plasmas in ST40. This is required for IR thermographic inversions leading to meaningful $\lambda_q$ estimates\cite{Eich_2013}. Additionally, the ST40 divertor geometry allows a strong optimisation of the diagnostic system, instead intrinsically more challenging in presence of a closed divertor with vertical targets \cite{CMOD_IR}.
        
    \subsection{Diagnostic apparatus: IR camera}\label{sec:ir_camera}

        \subsubsection{Hardware characteristics}\label{sec:ir_camera_hardware}

\begin{figure}
                \centering
                \subfloat[]{\includegraphics[height = 5.45 cm]{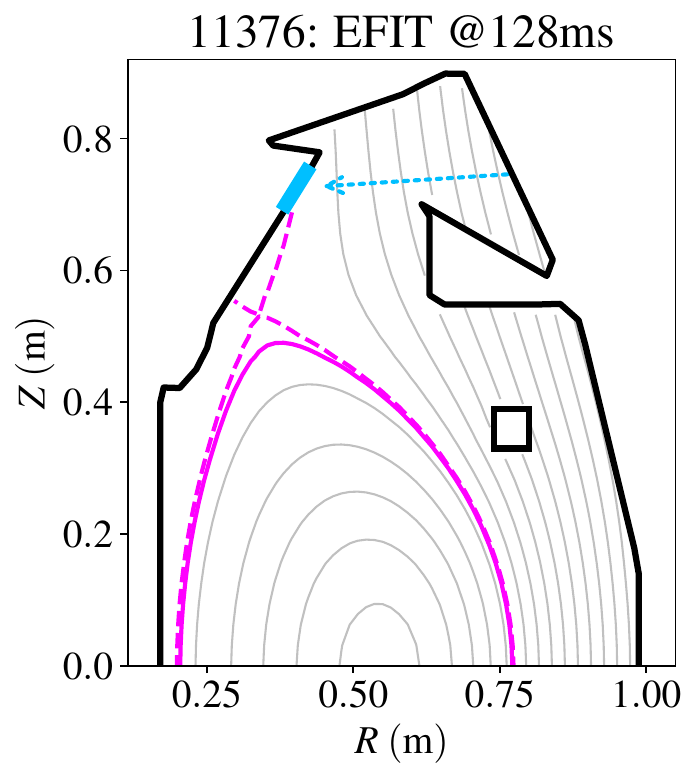}\label{fig:bundle-a}}
                \hspace{0.025 cm}
                \subfloat[]{\includegraphics[height = 5.45 cm]{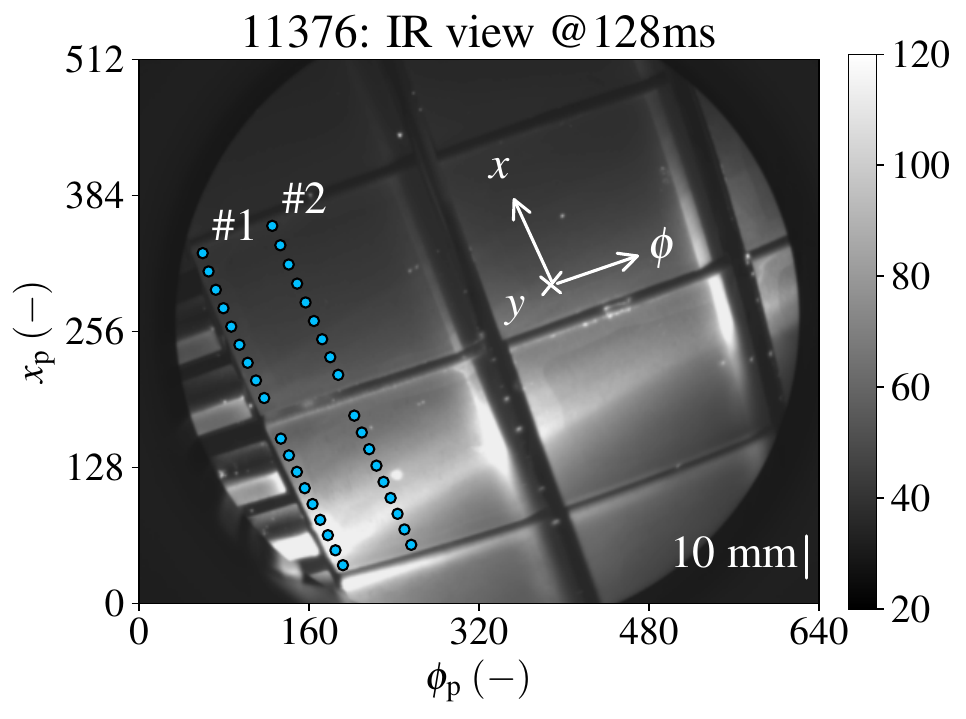}\label{fig:bundle-b}}
                \hspace{0.025 cm}
                \subfloat[]{\includegraphics[height = 5.45 cm]{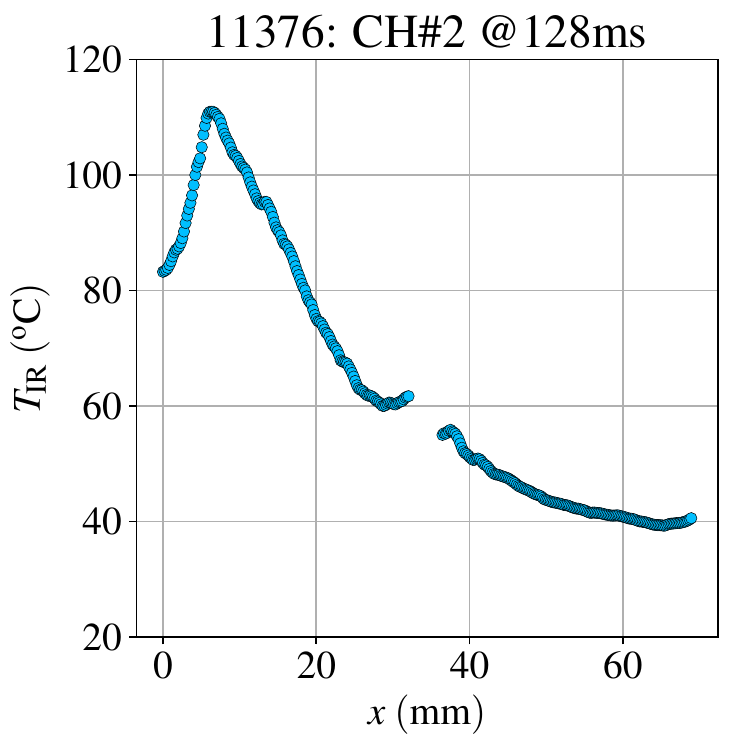}\label{fig:bundle-c}}
                \caption{Shot 11376 of the ST40 database at 128 ms. (a) EFIT reconstruction of the disconnected double-null magnetic equilibrium: primary and secondary separatrix in magenta solid and dashed, respectively; further magnetic surfaces in grey; wall in black; CH\#2 in blue (1:1 scale); LoS in dashed blue. (b) IR camera view on the upper outer divertor in pixel space $(x_{\text{p}},\phi_{\text{p}})$. Overlaid are (see text for details): the FAHF reference frame in real space with $\phi$, proxy for toroidal direction, $x$, proxy for poloidal direction, and the direction in the depth of the tiles $y$ (entering the screen); an illustrative representation of the chords CH\#1 and CH\#2 (blue dots) of the bundle in both their sub-chords, one per tile. (c) Temperature $T_{\text{IR}}$ as extracted from the IR view via the Calcam software \cite{calcam} (Appendix \ref{sec:calcam}) along CH\#2 in real space, with the empty gap in its middle. The data-point resolution is the actual $0.21 \pm 0.03$ mm/pixel of the ST40 IR camera.}  
                \label{fig:bundle}
            \end{figure}

            ST40 is equipped with a FLIR X6903sc MWIR ($3-5 \;\mu$m) camera, with an acquisition rate of 0.8 kHz ($\Delta t_{\text{IR}} = $ 1.25 ms) for its 612 $\times$ 540 pixels, and diagnoses the upper outer divertor target (Fig. \ref{fig:bundle-a}). The reader is redirected to Refs.\cite{ir}, \cite{ASDEX_IR_2} and \cite{Adebayo-Ige_2024} for the fundamental principles of IR thermography, and to Table \ref{tab:camera} for the specifications of the ST40 system, further commented below.
    
            \begin{table}
                \centering
                \caption{Specifications of ST40's FLIR X6903sc MWIR camera: diagnosed wavelength spectrum; acquisition rate; time resolution; pixels along $\phi_{\text{p}}$ and $x_{\text{p}}$ directions (Fig. \ref{fig:bundle-b}); average spatial resolution along the divertor; field of view (FoV) along $\phi_{\text{p}}$ and $x_{\text{p}}$ in degrees (deg); angle between the line of sight (LoS) and divertor surface normal; calibrated dynamic temperature range; saturation temperature.}
                \renewcommand*\arraystretch{1.4}
                \resizebox{\textwidth}{!}{
                \begin{tabular}{|c|c|c|c|c|c|c|c|c|}
                    \hline
                    Spectrum ($\mu$m) & Rate (kHz) & $\Delta t_{\text{IR}}$ (ms) & Pixels ($-$) & $\Delta x_{\text{IR}}$ (mm/pix) & FoV (deg) & LoS angle (deg) & Range (\textcelsius{}) & Saturation (\textcelsius{}) \\\hline %\hline
                    $3-5$ & 0.8 & 1.25 & 640 $\times$ 512 & 0.21 $\pm$ 0.03 & ($14.6 ; 12.7$) & 38 & [$12;106$] & 118 \\\hline
                \end{tabular}
                }
                \label{tab:camera}
            \end{table}

            \begin{figure}
                \centering
                % \subfloat[]{\includegraphics[height = 5.45 cm]{ST40_EFIT_128ms.pdf}\label{fig:bundle-a}}
                \includegraphics[width = 0.485\textwidth]{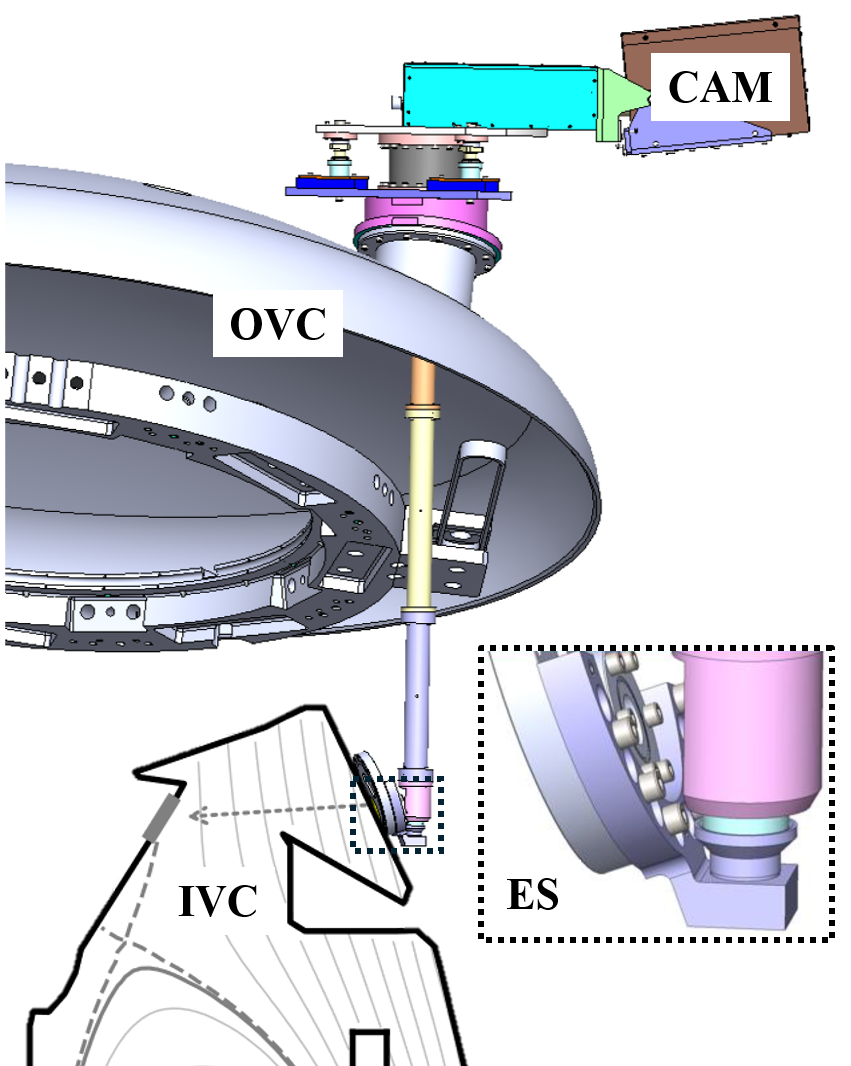}
                \caption{Main components of the endoscope assembly in ST40: camera protective housing (CAM); outer vacuum chamber (OVC); zoomed-in endoscope support (ES); schematic of the inner vacuum chamber (IVC), further detailed in Fig. \ref{fig:bundle-a}.}  
                \label{fig:hardware}
            \end{figure}
            
            An \textit{ad-hoc} re-entrant endoscope has been designed for ST40 with Zinc Selenide (ZnSe) optics, and is pictured in Fig. \ref{fig:hardware}. The horizontal sub-assembly (top) contains the camera protective housing (CAM) and is fixed to the vertical sub-assembly, penetrating the outer vacuum chamber (OVC), by means of a connection flange. The assembly terminates with endoscope support (ES), facing the inner vacuum chamber (IVC).
            \\The resulting camera line of sight (LoS, dashed blue line in Fig. \ref{fig:bundle-a}) within the IVC is at an acceptable angle\footnote{Data related to LoS, FoV and pupil position are extracted via Calcam\cite{calcam} (Appendix \ref{sec:calcam}).} of 38 degrees to the divertor surface normal (LoS angle). Instead, the field of view (FoV, 14.6 degrees along $\phi_{\text{p}}$ and 12.7 degrees along $x_{\text{p}}$ in Fig. \ref{fig:bundle-b}) is such that to cover the minimum amount of surface required to fully capture the heat flux footprint. The compenetration of the above LoS, FoV, short pupil-divertor distance (0.37 m) and the accommodating divertor geometry (Sec. \ref{sec:st40}) results in a noteworthy effective spatial resolution of the camera, averaging $\Delta x_{\text{IR}} = 0.21 \pm 0.03$ mm/pixel ($\sim 5$ pixel/mm) over two divertor tiles, and appreciable in Fig. \ref{fig:bundle-c}. The comparison with the list of references in Sec. \ref{sec:introduction} actually suggests the spatial resolution of ST40 IR camera to be the highest available worldwide for thermographic inversions of a full plasma heat flux footprint.
            \\Worthwhile stressing is also the orientation of the camera LoS relative to the plasma magnetic topology: the LoS prevalently crosses the optically-thin far SOL, and is thus less contaminated by plasma absorption/emission in the IR spectrum.

        \subsubsection{Calibration and surface properties}\label{sec:ir_camera_calib}
            
            From the original IR camera dynamic range $[10 ; 90]$ \textcelsius{}, factory-calibrated with its 50-mm lens, and the in-house characterisation of the endoscope transmittance yielding 0.85, the resulting calibrated range for this work is $[12 ; 106]$ \textcelsius{}. The calibration is still applicable in $(106 ; 118]$ \textcelsius{}, with an accuracy however diminishing as the IR camera saturation temperature of $118$ \textcelsius{} is approached.
            \\The surface emissivity of the divertor tiles is assumed to be 1.0, constant with respect to the position in space and surface temperature. Despite the molybdenum divertor surface, no spurious reflection is observed to affect the view.
            \\Neither the calibration factor/range nor the surface emissivity impact the \textit{numerical} characterisation of FAHF, focus of the present study. Also, Appendix \ref{sec:noise} suggests that only a small correction on the output should be expected from an uncertainty of these parameters. A dedicated assessment of such factors will be accomplished, being instead required for any \textit{physical} investigation of ST40 edge plasma.
            
   \subsection{Reference shot}\label{sec:ref_shot}
   
        Shot number 11376 within the ST40 database is selected for the analyses. According to database-wide analyses\cite{zhang2024experimental}, this is a representative ST40 plasma shot from the edge plasma point of view, and features a single-exponential heat flux footprint (Sec. \ref{sec:sensitivity_Eich}). The absence of edge-localised modes \cite{NSTX_ELMs} (ELMs) evolving on short time-scales simplifies the numerical investigations (Sec. \ref{sec:comparison_fahf_comsol2d_comsol3d}).
        \\The plasma presents a stable flat-top during the time window $[50 ; 140]$ ms, when the plasma is NBI-heated with $\sim 0.8$ MW of power. Although the IR camera is marginally affected by saturation in $[130;140]$ ms in the region of interest (Sec. \ref{sec:ref_chord}), this does not impact the numerical assessment of FAHF.
        \\The magnetic equilibrium of shot 11376 (Fig. \ref{fig:bundle-a}) is nominally a lower-biased disconnected double-null (Sec. \ref{sec:st40}), i.e. the primary separatrix contacts the lower divertor, thereupon rerouting most of the power\cite{Brunner_2018, Osawa_2023}. However, the low degree of disconnection ($\delta r_{\text{sep}} / \lambda_q \sim 1.5$ at the outer mid-plane \cite{Matteo_PSI}) allows the plasma to still deposit a satisfactory amount of power on the upper divertor (in view of the camera) during the actual divertor heating phase in $[90;140]\;\text{ms}$.

\section{Fundamentals of IR thermographic inversion}\label{sec:theory_general}

    \subsection{Solving the heat equation}\label{sec:heat_eq_theory}

        By safely discarding any volumetric heat source/sink\footnote{For instance, adding the volumetric Ohmic heating caused by the $\sim 1$ A of current drew by a Langmuir probe \cite{Hutchinson_2002} in ST40 does not cause any appreciable change in the output.}, the general form of the unsteady homogeneous heat conduction equation for an isotropic medium in cartesian coordinates is given by \cite{incropera}:
        \begin{equation}\label{eq:heat_eq}
            \varrho c_{\text{p}} \frac{\partial T}{\partial t} = \nabla \cdot (\kappa \nabla T) \:,
        \end{equation}
        with $\varrho$ being the mass density ($\text{kg} \times \text{m}^{-3}$), and $c_{\text{p}}$ the specific heat capacity at constant pressure ($\text{J} \times \text{kg}^{-1} \times \text{K}^{-1}$). Lastly, $\kappa$ is the heat conductivity ($\text{W} \times \text{m}^{-1} \times \text{K}^{-1}$), which is acted upon by the divergence operator.
        \\Upon imposing boundary conditions (BCs) in space, and an initial condition (IC) in time at $t = 0$ ms, Eq. (\ref{eq:heat_eq}) is solved for the temperature $T$. According to the methodology specific to the thermographic inversion, a Dirichlet-type BC is enforced via the experimental temperature data $T_{\text{IR}}$ (Fig. \ref{fig:bundle-c}) at the plasma-facing boundary of the modelled domain. On the remaining non-plasma-facing boundaries, homogeneous Neumann BCs are assumed, physically representing adiabatic (no-flux) boundaries. This is legitimate in ST40 because of a negligible black-body emission in the gap ($T_{\text{IR}} < 50$ \textcelsius{}), and of a negligible heat convection/conduction provided by the rarefied neutral gas filling this volume (few Pa in pressure at most \cite{Elena_PSI}). Additionally, the thermal contact of the tile back-surface is neglected. Applying various Dirichlet BCs at the back-surface is investigated, but does not cause any noticeable corrections in the time window $[90;140]$ ms, consistently with Sec. \ref{sec:theory} and \ref{sec:geometrical_error_time}.

    \subsection{Surface heat flux and Eich fit}\label{sec:heat_flux_theory}
        
        The temperature distribution $T$ from Sec. \ref{sec:heat_eq_theory} informs the computation of the quantity of primary physical interest of the IR thermographic inversion, the heat flux vector at the PFC surface $(\text{W} \times \text{m}^{-2})$:
        \begin{equation}\label{eq:heat_flux_theory}
            \mathbf{q} = -\kappa \nabla T \:, 
        \end{equation}
        and, specifically, its perpendicular component $\mathbf{q} \cdot \mathbf{n}$ (herein referred to as "surface heat flux") along the tile surface normal direction $\mathbf{n}$ (defined as entering in the tile). Once this quantity is calculated, the thermographic inversion is nominally concluded.
        \\However, further information about the SOL plasma is customarily retrieved by fitting the surface heat flux profile in space with the Eich function $q_{\text{Eich}}$ from Eq. (1) of Ref.\cite{Eich_2013}. The main output of such fitting procedure is embodied by the parameters of: maximum heat flux $q_0$, power fall-off length $\lambda_q$, spreading factor $S$, background heat flux $q_{\text{BG}}$, strike point location $x_0$, and poloidal flux expansion $f_X$. Ref.\cite{Eich_2013} is recommended for detailed descriptions of such quantities.
        \\The integral value of the Eich profile $q_{\text{Eich}}(x)$ along the divertor, proportional to the total power, is here additionally added, and defined as:
        \begin{equation}\label{eq:eich_integral}
            \text{Integral} \; (\text{W} \times \text{m}^{-1}) = \int_{-\infty}^{+\infty} q_{\text{Eich}}(x) \; dx \:.
        \end{equation}
        Although the formal iterative procedure of Ref.\cite{marsden2024} is actually best-posed when fitting with the multi-parameter non-linear function $q_{\text{Eich}}$, and should be more widely adopted in the community, this is not here accomplished because of its computational cost. Instead, the fit is deemed satisfactory upon showing $R^2 > 98.5 \%$, and is herein merely considered as a parametrisation of the surface heat flux, to extract insightful parameters to characterise FAHF's precision. This simplified approach is therefore acceptable, and is justified \textit{a posteriori} by Secs. \ref{sec:sensitivity_Eich} and \ref{sec:sensitivity_Eich_dy}, which show a genuine behaviour of the heat flux.
        \\The interested reader is redirected to Refs. \cite{marsden2024} and \cite{zhang2024experimental} for the actual quantification of $\lambda_q$ in ST40, which is outside the scope of the present work.

    \subsection{Internal energy variation and deposited energy}\label{sec:energies_theory}
    
        From Secs. \ref{sec:heat_eq_theory} and \ref{sec:heat_flux_theory}, the internal energy variation over a time interval $\tau$ throughout the volume $V$ of the domain:
        \begin{equation}\label{eq:energy_int_theory}
            \Delta\mathcal{E}_{\text{int}} = \int_{\tau} \int_{V} \varrho c_{\text{p}} \frac{\partial T}{\partial t} \times dV \: dt \:,
        \end{equation}
        and the net energy over $\tau$ flowing through the tile surface $\mathbf{S} = S \times \mathbf{n}$:
        \begin{equation}\label{eq:energy_surf_theory}
             \mathcal{E}_{\text{surf}} = \int_{\tau} \int_{S} \mathbf{q} \cdot \mathbf{n} \times dS \: dt
             \:,
        \end{equation}
        are computed. This aids the assessment of the energy-preserving properties of the algorithms. With $\mathbf{n}$ defined in Sec. \ref{sec:heat_flux_theory}, $\mathcal{E}_{\text{surf}} > 0$ when deposited in the tile.
        
\section{Numerical methods and tools}\label{sec:methods}

    \subsection{FAHF}\label{sec:fahf}

        The present section provides essential details of the assumptions, choices and implementation of FAHF algorithms.

         \subsubsection{Model assumptions}\label{sec:theory}

            Model assumptions are adopted in FAHF to simplify and speed up the solution of Eq. (\ref{eq:heat_eq}). Crucially, if $\kappa = \kappa(T)$ or it explicitly depends on space (e.g. multi-material PFCs), then $\nabla \cdot (\kappa \nabla T) = \nabla \kappa \cdot \nabla T + \kappa \nabla^2 T$, and a careful differentiation would be required. An elegant solution to this is employed in Ref.\cite{theodor_original}, however not needed in the specific case of ST40, where the divertor heating phase lasts $\sim 50$ ms at most ($[90;140]\;\text{ms}$, Sec. \ref{sec:ref_shot}) $-$ not enough to result in a temperature which would appreciably change the material properties. Therefore:
            \begin{itemize}
                \item Asm. \#1. The material properties are assumed to be temperature-independent.
            \end{itemize}
            No impurity surface layer \cite{Adebayo-Ige_2024, theodor} is accounted for. Moreover, the region affected by a temperature variation in a semi-infinite solid\footnote{With $L = 29$ mm of the present case, a reasonably constant heat flux over $[90;140]$ ms (Fig. \ref{fig:fahf_comsol2d_comsol3d}), and assuming $\text{Fo} = \alpha t / L^2 < 0.2$ \cite{incropera}, the approximation of semi-infinite solid would hold for up to $\sim 3$ s.} is here limited to $2 \sqrt{ \alpha \times 50 \;\text{ms}} \sim 3.3$ mm (with $\alpha \sim 54 \; \text{mm}^2 \times \text{s}^{-1}$ defined in Eq. (\ref{eq:heat_eq_fahf})) along the depth of the tile, hence still confined within the only Mo stratum. Consequently:
            \begin{itemize}
                \item Asm. \#2. The divertor tiles are assumed to be a single layer of Mo of 29 mm in thickness, instead of the actual 4 mm of plasma-facing Mo overlaid to 25 mm of CuCrZr.
            \end{itemize}
            Asms. \#1 and \#2 imply minor model errors in FAHF, as demonstrated in Sec. \ref{sec:comparison_fahf_comsol2d}, and Eq. (\ref{eq:heat_eq}) accordingly reduces to:     \begin{equation}\label{eq:heat_eq_fahf}
                \frac{\partial T}{\partial t} = \alpha \nabla^2 T \:,
            \end{equation}
            where $\alpha = \kappa / (\varrho c_{\text{p}})$ is the thermal diffusivity ($\text{m}^2 \times \text{s}^{-1}$), unaffected by any differential operator only by virtue of being assumed constant in the present case.

        \subsubsection{Dimensionality of the problem and geometrical assumption}\label{sec:dimensionality}
    
            The orthogonal reference frame used in FAHF is depicted in Fig. \ref{fig:bundle-b} and chosen as aligned with the sides of the cuboidal tiles. In particular, $x$ is the direction along the divertor (henceforth referred to as "poloidal direction"), $y$ the direction along the depth of the tile, and $\phi$ happens to be a proxy for the toroidal direction (which would actually be aligned to the strike point band). When solving Eq. (\ref{eq:heat_eq_fahf}) in FAHF:
            \begin{itemize}
                \item Asm. \#3. The two-dimensional description of heat conduction is assumed to be appropriate, and the toroidal direction is neglected by assuming toroidal symmetry.
            \end{itemize}
            A bundle of $N_{\phi}$ independent parallel observation chords is generated, two such being shown in Fig. \ref{fig:bundle-b} (blue dots), with a more complete representation found in Ref.\cite{marsden2024}. The chords are aligned with the $x$ direction and sit at uniformly-spaced locations $\{\phi_k\}_{k=1}^{N_{\phi}}$ on the rectangular planar surface (Sec. \ref{sec:st40}) of the divertor tile CAD ($x\phi$ plane). This CAD is a suitable representation of the actual tiles, as long as surface imperfections are absent. Surface intrusions/protrusions deterministically cause noticeable abnormalities in the results\cite{marsden2024}. Any chords affected is therefore discarded.
            \\Each chord is then further divided along $x$ in two equal sub-chords, one per tile, where the toridal-running gap lies. This is allowed by the thermal insulation provided by the gap (Sec. \ref{sec:BCs}). The resulting 2D computational domain in FAHF is therefore a (32 mm) $\times$ (29 mm) rectangle in the $xy$ plane. The overall output along the entire chord (e.g. surface heat flux profile, Sec. \ref{sec:heat_flux}) is ultimately obtained by juxtaposing in space the FAHF output of each of the two sub-chords at every time-step, as accomplished for $T_{\text{IR}}$ in Fig. \ref{fig:bundle-c}.
            \\Asm. \#3, resulting in FAHF being independently run along (each sub-chord of) each chord in a "multi-2D" fashion, is expected to be justified by the fact that the poloidal temperature gradients $\partial_{x} T_{\text{IR}}$ tends to dominate over the toroidal gradients $\partial_{\phi} T_{\text{IR}}$ in the majority of plasma properties. However, following from the peculiar three-dimensionality of the ST40 divertor, $\partial_{\phi} T_{\text{IR}} \neq 0$ still holds, as evident from Fig. \ref{fig:bundle-b}, and also detailed in Ref.\cite{marsden2024}. The geometrical error implied by the multi-2D approach, instead of a 3D one, is assessed in Sec. \ref{sec:comparison_fahf_comsol2d_comsol3d}, and shown to be within an acceptable limits in the bulk of the tile.
            \\Compared to the "single-chord" approach most commonly employed within the community, this multi-2D approach also beneficially provides statistics and relevant error-bars in the value of $\lambda_q$ ultimately estimated \cite{marsden2024}. Moreover, the multi-2D treatment properly accounts for difference of the perpendicular heat flux on a chord-to-chord basis, crucial for a power balance analysis, particularly in ST40.
    
        \subsubsection{Numerical discretisation in space and time}\label{sec:discretisation}
        
            The finite difference (FD) method is relied upon for the 2D spatial discretisation in the nodes of the rectangular domain, for reasons of simplicity and speed.
            \\In particular, $N_{x}$ points $\{x_i\}_{i=1}^{N_{{x}}}$ are placed along each sub-chord ($x$ direction), with uniform spacing $\Delta x = x_{i+1} - x_i \; \forall i \in [1 ; N_{x})$. Instead, $N_{y}$ points $\{y_j\}_{j=1}^{N_{{y}}}$ are distributed along the tile depth ($y$ direction), with uniform spacing $\Delta y = y_{j+1} - y_j \; \forall j \in [1 ; N_{y})$ (potentially differing from $\Delta x$), where $j=1$ identifies the plasma-facing surface, and $j=N_y$ the back-surface. The second-order centred FD scheme\cite{incropera} is then leveraged in the interior nodes $(i,j) \in [2 ; N_x-1] \times [2 ; N_y-1]$.
            \\Non-uniform grids along $y$ are found in the literature \cite{Adebayo-Ige_2024}, but are advantageous under specific circumstances outside the scope of the present work (e.g. when including an impurity surface layer\cite{hermann}). The space convergence analysis of Sec. \ref{sec:space_time_convergence} justifies the adoption of a uniform mesh.
            \\In time, the first-order explicit forward Euler time stepping scheme is implemented at the time instants $\{t^n\}_{n=1}^{N_t} \in [0;200]$ ms, with uniform time-step $\Delta t = t^{n} - t^{n-1} \; \forall n \in (1 ; N_t]$. This implementation must satisfy the stability criterion \cite{incropera}:
            \begin{equation}\label{eq:stability}
                \Delta t \times \left(\frac{1}{\Delta x^2} + \frac{1}{\Delta y^2} \right) < \frac{1}{2 \alpha} \:.
            \end{equation}
            Reasons behind selecting an explicit scheme are the ease of implementation and, with the proper user settings identified in Sec. \ref{sec:space_time_convergence}, a satisfactory speed. An implicit implementation would be considered, were it needed for specific applications\cite{Adebayo-Ige_2024} beyond those of the present work.
            \\The temperature at the node $(i,j)$ and time instant $n$ is finally denoted as $T_{i,j}^{n} = T(x_i,y_j,t^n)$.
            
        \subsubsection{Boundary and initial conditions}\label{sec:BCs}
        
            With the BC choices of Sec. \ref{sec:heat_eq_theory}, and exploiting the Calcam software\cite{calcam} according to Appendix \ref{sec:calcam}, the experimental datum $T_{\text{IR}}(x,\phi)$ is sampled over the set $\{x_i\}_{i=1}^{N_{x}}$ along each chord at every time instant $\{t^n\}_{n=1}^{N_{\text{t}}}$. The Dirichelet BC data $T_{i,1}^{n} = T_{\text{IR},i}^{n} \; \forall i,n$ is obtained, and enforced at the plasma-facing boundary ($j=1$). Although both the user-selected $\Delta x$ and the $\Delta t$ resulting form the stability criterion (Eq. (\ref{eq:stability})) might fall below the IR camera spatial and/or temporal resolution (Sec. \ref{sec:ir_camera_hardware}), this is not here of concern, as the linear interpolation of $T_{\text{IR}}$ is checked to not introduce any artifacts (e.g. over-/under-shooting). An assessment of the uncertainty on $T_{\text{IR}}$ is provided in Appendix \ref{sec:noise}.
            \\A first-order homogeneous Neumann BC is adopted for the non-plasma-facing boundaries ($\{i = 1\} \cup \{i=N_x\} \;\forall j>1$ or $j=N_y \;\forall i$), assumed adiabatic (Sec. \ref{sec:heat_eq_theory}).
            \\In the temporal domain, the IC $T_{i,j}^{1}$ chosen at the first time instant $n=1$ is a uniform distribution in the entire domain. Its value is the spatial-average temperature along the sub-chord as measured by the IR camera, i.e. $T_{i,j}^{1} = \sum_{i=1}^{N_x} T_{\text{IR},i}^{1} / N_x \; \forall i,j$. Different flavours of such IC are tested, not providing any appreciable difference in the outcome after a few time-steps.

        \subsubsection{Discretised form of the problem}\label{sec:problem}
            
            The resulting discretised form of the heat conduction problem FAHF solves throughout the entire plasma discharge ultimately reads:
            \begin{equation}\label{eq:heat_eq_fahf_discretised}
            \begin{split}
                & \begin{split}  
                    \frac{T_{i,j}^{n} - T_{i,j}^{n-1}}{\Delta t} & = \alpha \: \frac{T_{i+1,j}^{n-1} - 2 T_{i,j}^{n-1} + T_{i-1,j}^{n-1}}{\Delta x^2} + \\
                    & + \alpha \: \frac{T_{i,j+1}^{n-1} - 2 T_{i,j}^{n-1} + T_{i,j-1}^{n-1}}{\Delta y^2} + \\
                    & + o(\Delta x^2) + o(\Delta y^2) + o(\Delta t) \:;
                \end{split}\\\\
                & \begin{split}
                        & \text{IC: } T^1_{i,j} = \sum_{i=1}^{N_x} T_{\text{IR},i}^1 / N_x \text{ if $n=1 \; \forall i,j$} \:;\\
                        & \text{BC: } \text{Dirichlet } T^{n}_{i,1} = T_{\text{IR},i}^{n} \text{ if $j = 1 \; \forall i,n \geq 2$} \:;\\
                    & \begin{split}
                        \text{BC: } & \text{homogeneous Neumann (no-flux)} \\
                        & \text{if $\{i = 1\} \cup \{i=N_x\} \;\forall j>1,n \geq 2$} \\
                        & \text{or $j=N_y \;\forall i,n \geq 2$} \:.
                    \end{split}    
                \end{split}
            \end{split}
            \end{equation}
            The only unknown $T_{i,j}^{n} \; \forall n \in [2;N_t]$ is explicitly retrieved at each timestep, and embodies the output of the simulation. While the mathematical properties of the discretisation error $o(\Delta x^2) + o(\Delta y^2) + o(\Delta t)$ are listed in Appendix \ref{sec:algebra_energy}, its behaviour in FAHF is thoroughly scrutinised throughout Sec. \ref{sec:verification}, and shown to negligibly impact the main conclusions of the work.

        \subsubsection{Surface heat flux discretisation}\label{sec:heat_flux}
        
            The surface heat flux of Sec. \ref{sec:heat_eq_theory} is calculated in post-processing at the plasma-facing surface ($j=1$) according to the first-order discretisation:
            \begin{equation}\label{eq:q_1st}
                \begin{split}
                    q_{i}^{n} = & - \kappa \: \frac{T_{i,2}^{n} - T_{i,1}^{n}}{\Delta y} + \\
                    & + o(\Delta x^2) + o(\Delta y) + o(\Delta t) \:,
                \end{split}
            \end{equation}
            with $q_i^n = q_{i,1}^{n}$. The error $o(\Delta x^2) + o(\Delta t)$ is inherited from $T_{i,j}^{n}$ of Eq. (\ref{eq:heat_eq_fahf_discretised}), while $o(\Delta y^2) \ll o(\Delta y)$ by definition (Eq. \ref{eq:errors_properties}). Higher-order spatial discretisations\cite{theodor, ASDEX_IR_2} could be implemented, but are here not required, as proven in Secs. \ref{sec:energy_balance} and \ref{sec:sensitivity_Eich}.
    
        \subsubsection{Discretisation of internal energy variation and deposited energy}\label{sec:energies}
        
            The discretisation of Eqs. \ref{eq:energy_int_theory} and \ref{eq:energy_surf_theory} over one time-step ($\tau = [t ; t+\Delta t]$) yields:
            \begin{equation}\label{eq:energy_int}
                \begin{split}
                    \Delta\mathcal{E}_{\text{int}}^{n} & = \sum_{i=1}^{N_x} \sum_{j=1}^{N_y} \varrho c_{\text{p}} (T_{i,j}^{n} - T_{i,j}^{n-1}) \times \Delta x \Delta y + \\
                    & + o(\Delta x^2) + o(\Delta y^2) + o(\Delta t) \:,
                \end{split}
            \end{equation}
            for internal energy variation, and:
            \begin{equation}\label{eq:energy_surf}
                 \begin{split}
                     \mathcal{E}_{\text{surf}}^{n} & = \sum_{i=1}^{N_x} q_{i}^{n} \times \Delta x \Delta t + \\
                     & + o(\Delta x^2) + o(\Delta y) + o(\Delta t) \:,
                 \end{split}
            \end{equation}
            for the net deposited energy, respectively. This enables the characterisation of the energy balance in FAHF from a purely numerical standpoint, which ensures further trust in the choices of Secs. \ref{sec:discretisation} and \ref{sec:BCs}. In Eqs. (\ref{eq:energy_int}) and (\ref{eq:energy_surf}) the errors follow from Eqs. (\ref{eq:heat_eq_fahf_discretised}) and (\ref{eq:q_1st}), respectively.
            
        \subsection{COMSOL Multiphysics\textsuperscript{\textregistered}}\label{sec:comsol}

            \subsubsection{Overview and motivation}\label{sec:comsol_overview}
            
                COMSOL Multiphysics\textsuperscript{\textregistered}, herein abbreviated to "COMSOL", is a widely-adopted state-of-the-art tool for multiphysical modelling\cite{COMSOL}. This suite is here configured as a pure heat conduction solver, and utilises the finite element method (FEM) in space with quadratic Lagrange elements, and second-order backward time stepping \cite{COMSOL}.
                \\Further detailed in Secs. \ref{sec:comparison_fahf_comsol2d_comsol3d} and \ref{sec:comparison_fahf_comsol2d}, different geometries (2D and 3D) and physical models (w/ and w/o temperature-dependent properties, w/ and w/o CuCrZr) are employed in COMSOL. This is an attempt of quantifying the model/geometry errors of FAHF, ultimately resulting in an estimate of FAHF's accuracy within its simplifying Asms. \#1, \#2 and \#3.

            \subsubsection{Setup and limitations of the comparison}\label{sec:comsol_setup}
            
                Because of reasons related to ingent computational cost, only one divertor tile (bottom left in Fig. \ref{fig:bundle-b}) is modelled by COMSOL in 3D and, for consistency, only the corresponding sub-chord belonging to such tile is modelled in 2D. The rectangular domain matching FAHF's (Sec. \ref{sec:dimensionality}) is uniformly meshed in COMSOL with $N_x \times N_y = 100 \times 100$ ($N_{\phi} = 100$ in 3D) rectangular elements ($\Delta x = 3.5 \times 10^{-4}$ m and $\Delta y = 2.9 \times 10^{-4}$ m), and the same initial and boundary conditions (Sec. \ref{sec:BCs}) are adopted, with $\Delta t = 1 \times 10^{-4}$ s. Ancillary convergence tests (not shown) have been performed, and confirm that a negligible discretisation error affects the COMSOL results.
                \\However, the fundamental numerics differing in FAHF and COMSOL (FDM vs. FEM, first- vs. second-order time stepping) constitutes a spurious source of disagreement of unknown magnitude. For this reason, the entire Sec. \ref{sec:geometrical_error} is devoted to an in-depth analysis, to carefully motivate the disagreement of the code predictions.
                \\Noteworthy is also the need of de-activating the automatic smoothing of the resulting heat flux in COMSOL, which would otherwise constitute a further unwanted source of error, for not being implemented in FAHF.

    \subsection{Definitions and conventions}

        \subsubsection{Reference chords}\label{sec:ref_chord}
            From the chord bundle generated for the multi-2D analysis (Sec. \ref{sec:dimensionality}), the two chords CH\#1 and CH\#2 are taken as references for the assessment of Asm. \#3 (Sec. \ref{sec:comparison_fahf_comsol2d_comsol3d}), and are depicted in Fig. \ref{fig:bundle-b} (blue dots). These chords sit 1.0 mm and 12.5 mm away from the left tile edge along $\phi$, respectively, and are affected neither by magnetic shadowing \cite{marsden2024} nor by surface imperfections (Sec. \ref{sec:dimensionality}).
            \\If not specified otherwise, CH\#2 is implicitly referred to throughout the paper. The output of a complete multi-chord analysis is reported by Ref.\cite{marsden2024}.

        \subsubsection{Convergence error}\label{sec:convergence_error}
            
            The behaviour of the errors in time $\varepsilon_*(\Delta t)$ and space $\varepsilon_*(\Delta x)$ is assessed via time and space convergence, the latter forcing $\Delta x = \Delta y$ for the sake of simplicity. These procedures are a numerical characterisation of the precision of the algorithm (verification), ensure trust in the internal workings of the code, and suggest the most suitable choice of $\Delta t$ and $\Delta x$, within the limit posed by Eq. (\ref{eq:stability}). Specific details of these procedures are reported in Appendix \ref{sec:convergence_details}, while Tables \ref{tab:time_convergence} and \ref{tab:space_convergence} condense the actual $\Delta t$ and $\Delta x$ employed.
            \\Given the reference point $(x_{*}, y_{*}, t_*)$ defined in Appendix \ref{sec:convergence_details} with $t_* = 128$ ms, the reference temperature $\tilde{T}_*$ at such point is defined as the result obtained with the smallest time/space-step of the time/space convergence, i.e. the most precise instance, and used to compare the other cases against. The relative difference of the resulting $T_* = T(x_{*},y_{*}, t_*)$ of a certain simulation with respect to the reference is labelled as "convergence error" and it is computed according to:
            \begin{equation}\label{eq:error_convergence}
                \varepsilon_* = \frac{\left| T_{*} - \tilde{T}_{*} \right|}{\left| \tilde{T}_{*} \right|} \:.
            \end{equation}

        \subsubsection{Heat flux error}\label{sec:heat_flux_error}
            
            The relative difference of the resulting surface heat flux $q_i^n$ (Sec. \ref{sec:heat_flux}) of a given simulation (e.g. COMSOL, Sec. \ref{sec:comsol}) with respect to FAHF's $\tilde{q}_i^n$ (unless specified otherwise) is referred to as "heat flux error" $\varepsilon_q$. Such error leverages the $\ell^{1}$ norm, with the heat fluxes being interpolated in time and space onto the grid of the most refined case between the two.
            \\The error can be computed either via integration in space:
            \begin{equation}\label{eq:l1_norm_time}
                \varepsilon^{n}(q) = \frac{\sum_{i=1}^{N_x} \left| q_{i}^{n} - \tilde{q}_{i}^{n}\right|}{\sum_{x=1}^{N_x} \left|\tilde{q}_{i}^{n}\right|} \:,
            \end{equation}
            returning an error profile in time or, viceversa, via integration in time:
            \begin{equation}\label{eq:l1_norm_space}
                \varepsilon_{i}(q) = \frac{\sum_{n=1}^{N_t} \left| q_{i}^{n} - \tilde{q}_{i}^{n}\right|}{\sum_{n=1}^{N_t} \left|\tilde{q}_{i}^{n}\right|} \:,
            \end{equation}
            returning an error profile in space along $x$ (at the plasma-facing surface).
            \\This "integral" approach conveniently condenses all the areas of agreement/disagreement among two cases. Its only shortcoming is the over-estimation of the actual error if a one specific time interval or space region is of interest.

        \subsubsection{Energy balance error}\label{sec:energy_balance_error}

            Starting from Sec. \ref{sec:energies}, the assessment of the energy conservation in each case of time and space convergence is accomplished. Similarly to Sec. \ref{sec:convergence_error}, only the time instant $t_* = 128$ ms is considered in the error estimation.
            \\For a perfect energy balance to be satisfied, the internal energy variation of Eq. (\ref{eq:energy_int}) must match the net deposited energy of Eq. (\ref{eq:energy_surf}). The relative difference among $\Delta\mathcal{E}_{\text{int*}} = \Delta\mathcal{E}_{\text{int}}(t_*)$ and $\mathcal{E}_{\text{surf*}} = \mathcal{E}_{\text{surf}}(t_*)$ for each case, labelled as "energy balance error", is thus defined by:
            \begin{equation}\label{eq:error_energy_time}
                \varepsilon_*(\mathcal{E}) = \frac{\left| \Delta\mathcal{E}_{\text{int*}} - \mathcal{E}_{\text{surf*}} \right|}{\left| \Delta\mathcal{E}_{\text{int*}} + \mathcal{E}_{\text{surf*}} \right| / 2} \:.
            \end{equation}
            Because of the lack of a well-defined reference among the two quantities, their average $\left| \Delta\mathcal{E}_{\text{int*}} + \mathcal{E}_{\text{surf*}} \right| / 2$ is used as normalisation factor \cite{Moscheni_2022}.

%%%%%%%%%%%%%%%%%%%%%%%%%%%%%%%%%%%%%%%%%%%%%%%%%%%%%%%%%%
%%%%%%%%%%%%%%%%%%%%%%%%%%%%%%%%%%%%%%%%%%%%%%%%%%%%%%%%%%

\section{Results}\label{sec:results}

    \subsection{Quantification of the precision of FAHF}\label{sec:verification}

        Any results discussed in this section is computed along CH\#2 of Sec. \ref{sec:ref_chord}.

        \subsubsection{Time and space convergence}\label{sec:space_time_convergence}

\begin{figure}
            
                \centering
                \subfloat[]{
                \includegraphics[width = 0.485\textwidth]{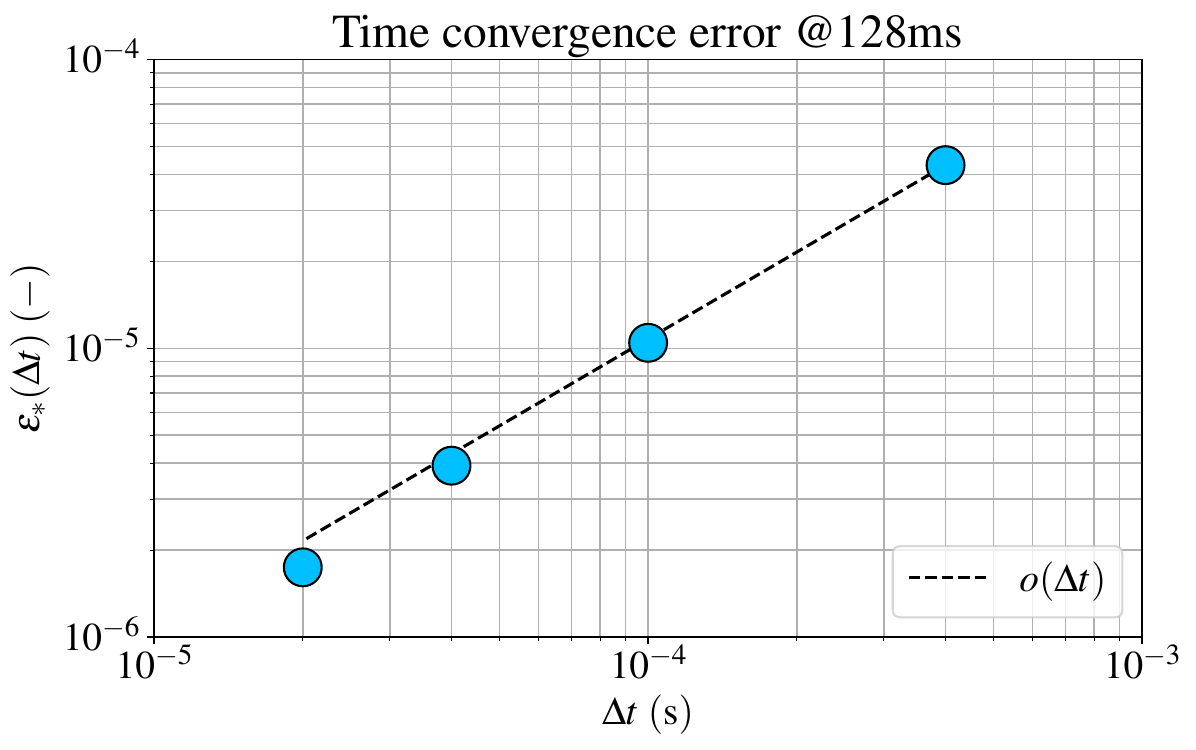}\label{fig:space_time_convergence-a}}
                \hspace{0.15 cm}
                \subfloat[]{\includegraphics[width = 0.485\textwidth]{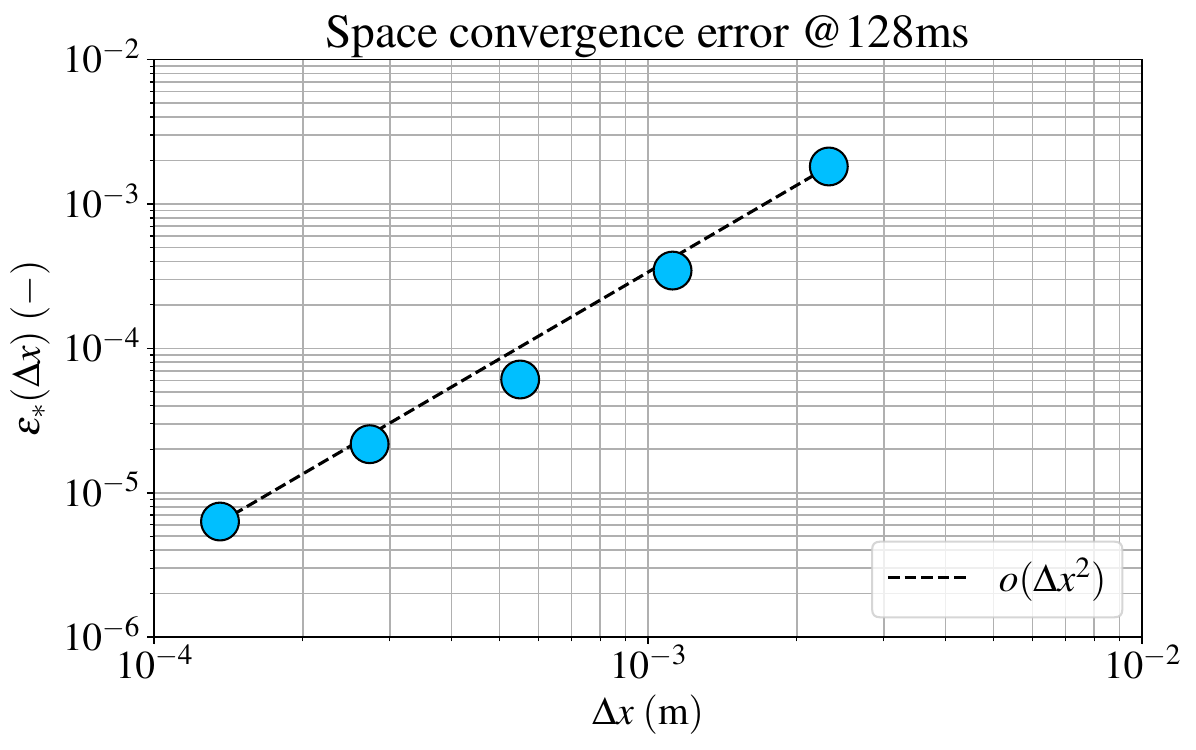}\label{fig:space_time_convergence-b}}
                \caption{(a) Time convergence and (b) space convergence of FAHF, showing the first-order and second-order decrease of the error evaluated as a function of $\Delta t$ and $\Delta x$, respectively.}  
                \label{fig:space_time_convergence}

            \end{figure}

            Following the guidelines of Sec. \ref{sec:convergence_error}, time and space convergence are assessed. Tables \ref{tab:time_convergence} and \ref{tab:space_convergence} respectively summarise the highlights of the two procedures, while Fig. \ref{fig:space_time_convergence} pictures the error behaviour.

            \begin{table}

                \centering
                \caption{Input parameters of the time convergence procedure (blue font for those which vary), and resulting errors.}
                \renewcommand*\arraystretch{1.4}
                \resizebox{0.5 \textwidth}{!}{
                \begin{tabular}{|c|c|c|c||c|c|}
                    \hline
                    $\Delta x \; (\text{m})$ & $N_x \; (-)$ & $\Delta t \; (\text{s})$ & $N_t \; (-)$ & $\varepsilon_*(\Delta t) \; (-)$ & $\varepsilon_*(\mathcal{E}) \; (-)$ \\\hline\hline
                    5.5E$-$04 & 60 & \textcolor{blue}{4.0E$-$04} & \textcolor{blue}{500}   & 4.3E$-$05 & 3.1E$-$01 \\\hline
                    5.5E$-$04 & 60 & \textcolor{blue}{1.0E$-$04} & \textcolor{blue}{2000}  & 1.0E$-$05 & 3.0E$-$01 \\\hline
                    5.5E$-$04 & 60 & \textcolor{blue}{4.0E$-$05} & \textcolor{blue}{5000}  & 3.9E$-$06 & 3.0E$-$01 \\\hline
                    5.5E$-$04 & 60 & \textcolor{blue}{2.0E$-$05} & \textcolor{blue}{10000} & 1.7E$-$06 & 3.0E$-$01 \\\hline
                    5.5E$-$04 & 60 & \textcolor{blue}{4.0E$-$06} & \textcolor{blue}{50000} & $-$ & 2.9E$-$01 \\\hline
                \end{tabular}
                }
                \label{tab:time_convergence}

                \centering
                \caption{Input parameters of the space convergence procedure (blue font for those which vary), and resulting errors. Reference FAHF simulation in boldface.}
                \renewcommand*\arraystretch{1.4}
                \resizebox{0.5 \textwidth}{!}{
                \begin{tabular}{|c|c|c|c||c|c|}
                    \hline
                    $\Delta x \; (\text{m})$ & $N_x \; (-)$ & $\Delta t \; (\text{s})$ & $N_t \; (-)$ & $\varepsilon_*(\Delta x) \; (-)$ & $\varepsilon_*(\mathcal{E}) \; (-)$ \\\hline\hline
                    \textcolor{blue}{2.3E$-$03} & \textcolor{blue}{15}  & 7.2E$-$06 & 27777 & 2.1E$-$03 & 1.1E$-$00 \\\hline
                    \textcolor{blue}{1.1E$-$03} & \textcolor{blue}{30}  & 7.2E$-$06 & 27777 & 4.0E$-$04 & 5.6E$-$01 \\\hline
                    \textcolor{blue}{5.5E$-$04} & \textcolor{blue}{60}  & 7.2E$-$06 & 27777 & 7.1E$-$05 & 2.5E$-$01 \\\hline
                    \textcolor{blue}{2.7E$-$04} & \textcolor{blue}{120} & 7.2E$-$06 & 27777 & 2.5E$-$05 & 1.2E$-$01 \\\hline
                    \textbf{\textcolor{blue}{1.4E$-$04}} & \textbf{\textcolor{blue}{240}} & \textbf{7.2E$-$06} & \textbf{27777} & \textbf{7.1E$-$06} & \textbf{5.7E$-$02} \\\hline
                    \textcolor{blue}{6.8E$-$05} & \textcolor{blue}{480} & 7.2E$-$06 & 27777 & $-$ & 2.8E$-$02 \\\hline
                \end{tabular}
                }
                \label{tab:space_convergence}

            \end{table}

            In particular, Fig. \ref{fig:space_time_convergence-a} shows how the ideal first-order convergence (black dashed line) is closely achieved by the error points $\varepsilon_*(\Delta t)$. Similarly, Fig. \ref{fig:space_time_convergence-b} confirms the second-order convergence of the error in space $\varepsilon_*(\Delta x)$, hence verifying the space convergence of the algorithms. The minor departures of the data-points from the ideal trend (e.g. $\Delta x = 5.5 \times 10^{-4}$ m in Fig. \ref{fig:space_time_convergence-b}) are not of concern, and could be ascribed to a combination of errors arising from interpolation and boundary conditions (Sec. \ref{sec:BCs}).
            \\In conclusion, the combination $\Delta t = 7.2 \times 10^{-6}$ s and $\Delta x = \Delta y = 1.4 \times 10^{-4}$ m (boldface in Table \ref{tab:space_convergence}) is chosen as a reference for further analyses (Sec. \ref{sec:comparison}). This choice guarantees a satisfactory precision in $T_{i,j}^n$, at an affordable computational cost of 158 s per sub-chord, scaling as $\propto N_x \times N_y \times N_t$. Notably, if the time step was instead chosen according to Eq. (\ref{eq:stability}), i.e. $\Delta t = 3 \times 10^{-5}$ s, the computational time would drop to 35 s per sub-chord.

        \subsubsection{Energy balance}\label{sec:energy_balance}
        
\begin{figure}
                \centering
                \includegraphics[width = 0.485\textwidth]{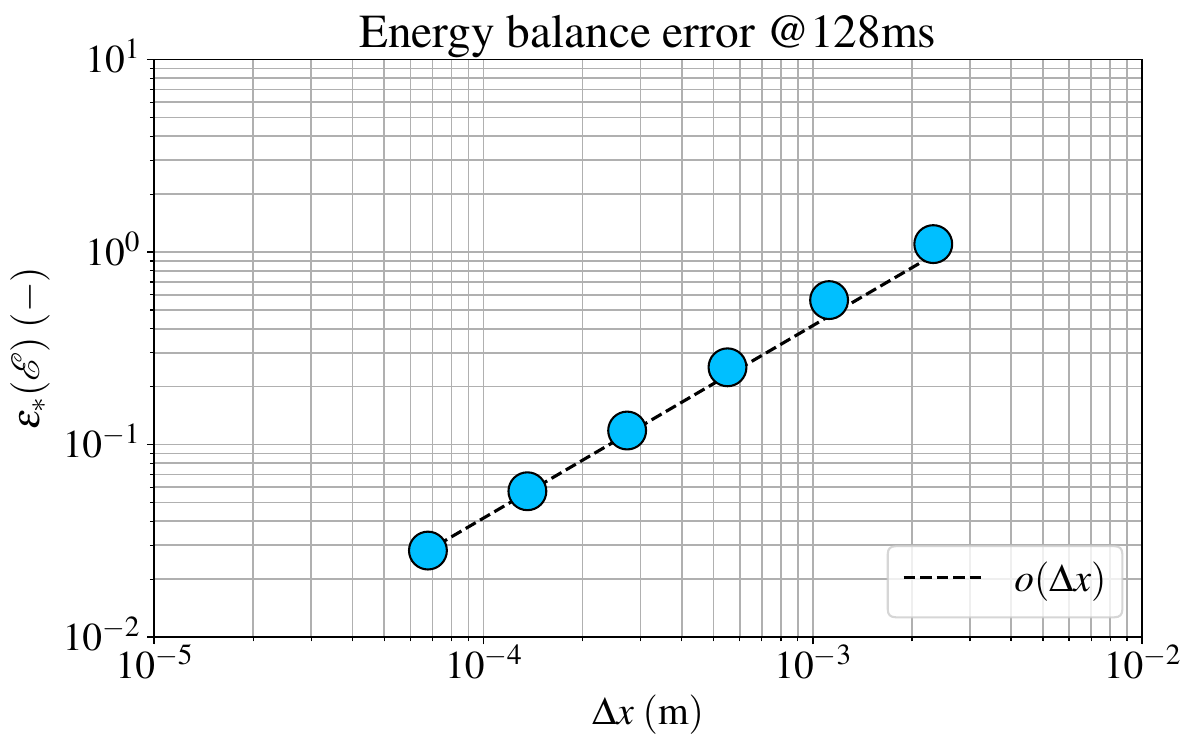}
                \caption{Energy balance error $\varepsilon_*(\mathcal{E})$ as a function of $\Delta x$.}
                \label{fig:E_balance}
            \end{figure}

            Fig. \ref{fig:E_balance} depicts the behaviour of $\varepsilon_*(\mathcal{E})$ as a function of $\Delta x$ (Sec. \ref{sec:energy_balance_error}), following from the data of which in Table \ref{tab:space_convergence} (last column). The error $\varepsilon_*(\mathcal{E})$ is found to show a first-order convergence with $\Delta x = \Delta y$, and almost no dependence on $\Delta t$ (Table \ref{tab:time_convergence}, last column). This is justified in Sec. \ref{sec:energy_balance_convergence}.
            \\The error $\varepsilon_*(\mathcal{E}) = 5.7\%$ achieved according to Table \ref{tab:space_convergence} (boldface) with the chosen $\Delta t = 7.2 \times 10^{-6}$ s and $\Delta x = \Delta y = 1.4 \times 10^{-4}$ m (Sec. \ref{sec:space_time_convergence}) is suitable for the current analyses, also because of the uncertainties customarily affecting experimental investigations of energy balance in tokamaks \cite{Matthews_2017, Redl_2023}, at least of the same order of magnitude.
            \\It is further noted that the actual sign\footnote{Recovered by removing the absolute value at the numerator of Eq. (\ref{eq:error_energy_time}).} of $\varepsilon_*(\mathcal{E})$ is such that $\Delta \mathcal{E}_{\text{int}}(t) > \mathcal{E}_{\text{surf}}(t) \; \forall t \in [40;140]$ ms if $\Delta x \geq 5.5 \times 10^{-4}$ m. Instead, no well-defined sign of the error is present in the high-resolution cases $\Delta x \leq 2.7 \times 10^{-4}$ m.
            
    \subsubsection{Surface heat flux: $\Delta x = \Delta y$}\label{sec:sensitivity_Eich}

        The surface heat flux resulting from the set of simulations of Sec. \ref{sec:space_time_convergence} (with $\Delta x = \Delta y$) is analysed according to Sec. \ref{sec:heat_flux}, at time instant $t_* = 128$ ms. As found for the energy error in Sec. \ref{sec:energy_balance}, $q_i^n$ does also not show a dependence on $\Delta t$, with the Eich fit parameters varying by 2\% at most. Therefore, only the results from the simulations of space convergence in Table \ref{tab:space_convergence} are detailed in Table \ref{tab:space_convergence_eich}. Full $q_i^n$ profiles in space are then shown by the dots in Fig. \ref{fig:q_space-a}, with their corresponding fitted Eich functions (solid lines). The main fitting parameters are pictured in Fig. \ref{fig:q_space-b}.
        \\On the one hand, a strong dependence on $\Delta x = \Delta y$ is observed, with the main fitting parameters in fact changing by up to a factor 2.5 as the resolution increases. On the other hand, a convergence of both the heat flux profiles and the corresponding fitting parameters is successfully achieved as $\Delta x = \Delta y$ drops (magenta to cyan). Any imperfect convergence is ascribed to the simplified fitting procedure here adopted (Sec. \ref{sec:heat_flux}). 
        \\The present section thus justifies how $\Delta t = 7.2 \times 10^{-6}$ s and $\Delta x = \Delta y = 1.4 \times 10^{-4}$ m (Sec. \ref{sec:space_time_convergence}) are also suitable choices from the perspective of $q_i^n$, which is in close reach of the converged value (boldface in Table \ref{tab:space_convergence_eich}).

    \subsubsection{Surface heat flux: $\Delta x \neq \Delta y$}\label{sec:sensitivity_Eich_dy}

\begin{figure}
            \centering
            \subfloat[]{\includegraphics[width = 0.485\textwidth]{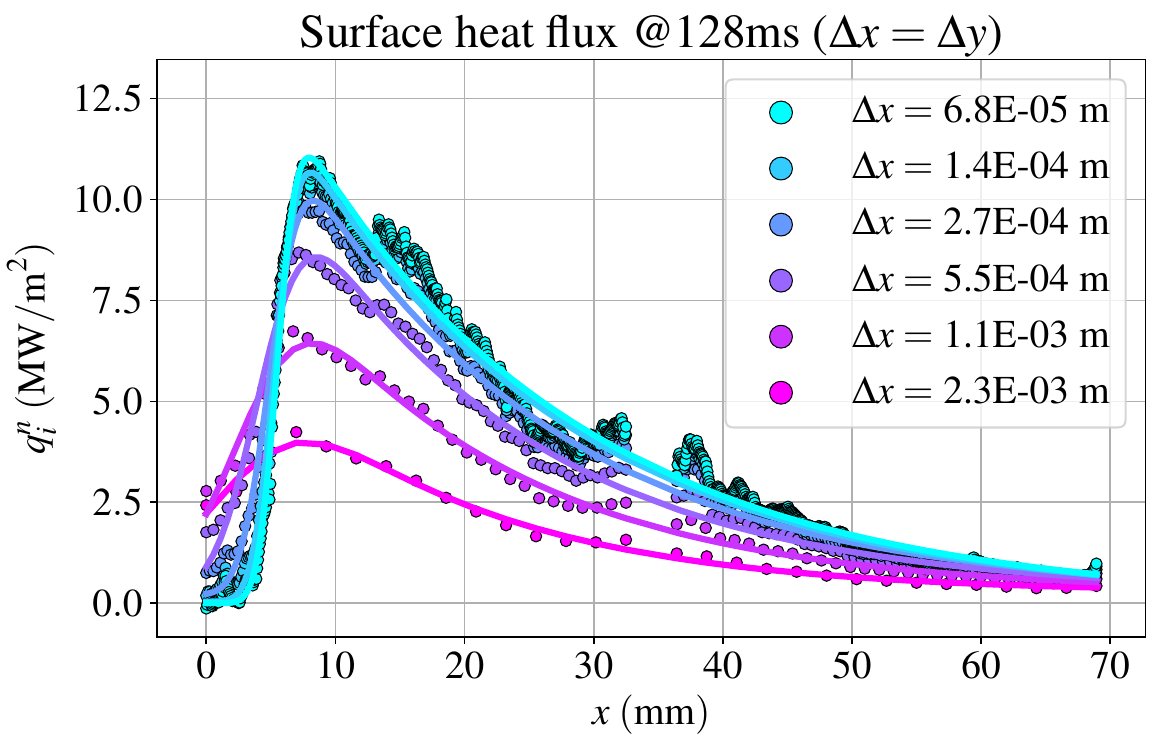}\label{fig:q_space-a}}
            \hspace{0.15 cm}
            \subfloat[]{\includegraphics[width = 0.485\textwidth]{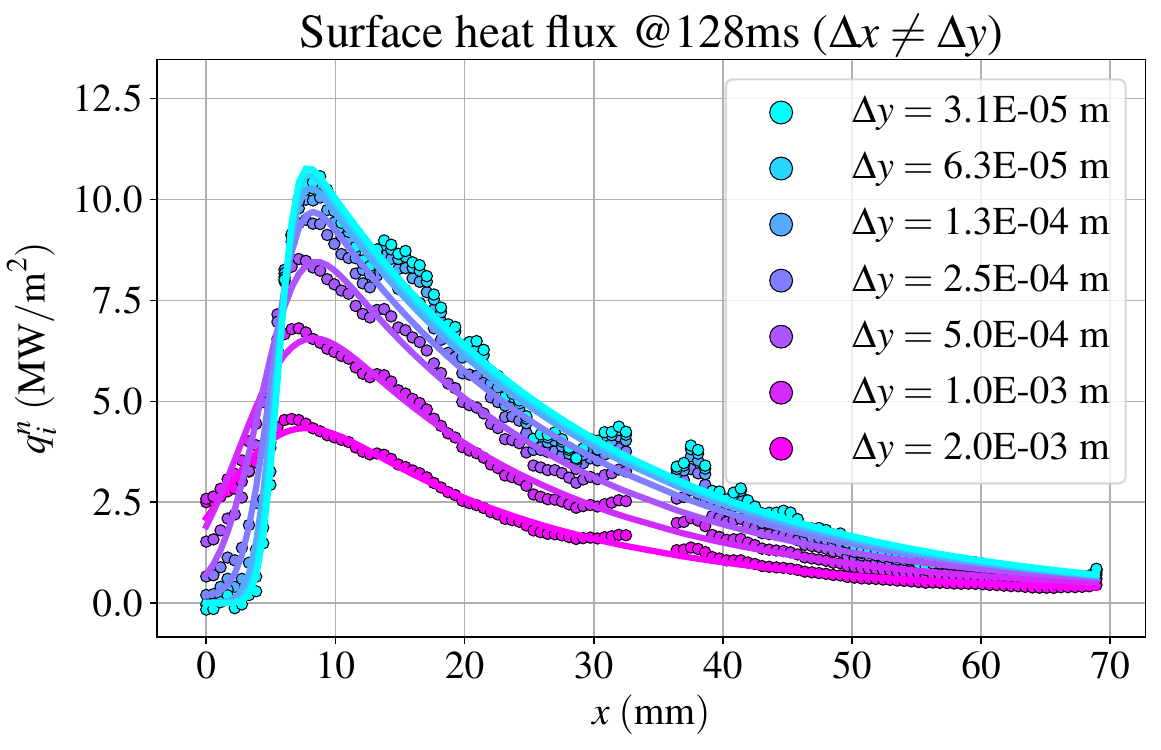}\label{fig:q_space_dy-a}}
            \\
            \subfloat[]{\includegraphics[width = 0.485\textwidth]{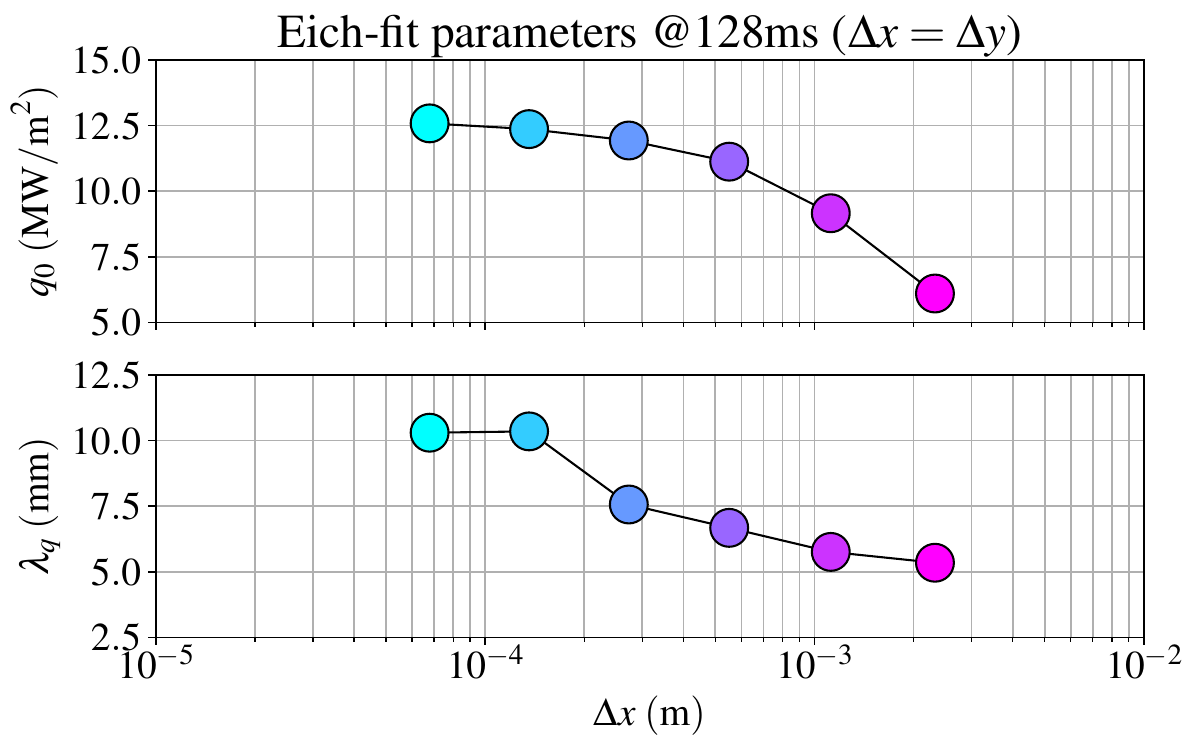}\label{fig:q_space-b}}
            \hspace{0.15 cm}
            \subfloat[]{\includegraphics[width = 0.485\textwidth]{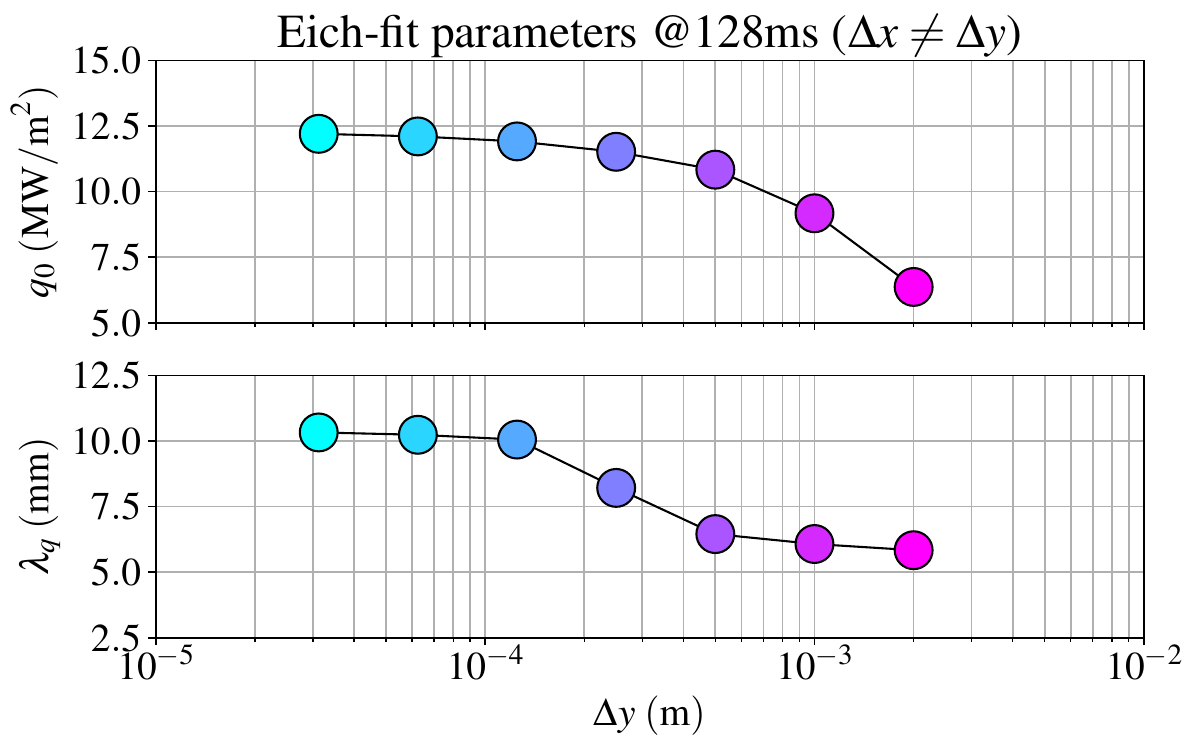}\label{fig:q_space_dy-b}}
            \caption{Surface heat flux profiles computed by FAHF (dots), in the cases of (a) Table \ref{tab:space_convergence} with $\Delta x = \Delta y$ and (b) Table \ref{tab:space_convergence_eich_dy} with $\Delta y \neq \Delta x$. Corresponding Eich fits in solid lines ($R^2 > 98.5\%$). Main Eich fit parameters as a function of (c) $\Delta x$ and (d) $\Delta y$.} 
            \label{fig:q_space}
        \end{figure}

        For Eq. (\ref{eq:q_1st}) depends on $\Delta y$ to the leading order, $\Delta x = 5.5 \times 10^{-4}$ m is here kept fixed, with $\Delta t = 7.2 \times 10^{-6}$ s (Table \ref{tab:space_convergence}). Only $\Delta y$ is scanned according to Table \ref{tab:space_convergence_eich_dy} (first column), table which then reports the entire outcome of the scan at time $t_* = 128$ ms. Heat flux profiles in space and main Eich fitting parameters are then depicted in Fig. \ref{fig:q_space_dy-a} and \ref{fig:q_space_dy-b}, respectively.
        \\The fitting parameters successfully converge towards fixed values, within a satisfactory 8\% of their counterparts in Table \ref{tab:space_convergence_eich} ($\Delta x = \Delta y$). This is however achieved at a diminished computational cost ($\propto N_x \times N_y \times N_t$): for instance, running FAHF with the $\Delta x = 5.5 \times 10^{-4}$ m and $\Delta y = 1.3 \times 10^{-4}$ m ($N_x = 60$ and $N_y = 232$) of Table \ref{tab:space_convergence_eich_dy} reduces the computational time of a factor $3.6$ compared to the reference $\Delta x = \Delta y = 1.4 \times 10^{-4}$ m ($N_x = 240$ and $N_y = 207$), for the same $\Delta t = 7.2 \times 10^{-6}$ s. If $\Delta t$ was instead independently chosen according to Eq. (\ref{eq:stability}), $q_0$ and $\lambda_q$ matched within 5\% could be achieved with a factor 6 worth of save in computational time: $\Delta t = 9.1 \times 10^{-5}$ s ($N_t = 2223$) for $\Delta x = \Delta y = 1.4 \times 10^{-4}$ m, and $\Delta t = 1.5 \times 10^{-4}$ s ($N_t = 1334$).

        \begin{table}
        
            \centering
            \caption{Eich fitting parameters \cite{Eich_2013} and Eich integral (Eq. (\ref{eq:eich_integral})) as a function of $\Delta x = \Delta y$ in the simulated cases of Table \ref{tab:space_convergence}. Reference FAHF simulation in boldface (see text for details).}
            \renewcommand*\arraystretch{1.4}
            \resizebox{\textwidth}{!}{
            \begin{tabular}{|c|c|c|c||c|c|c|c|c|c|c|}
                \hline
                $\Delta x$ & $N_x$ & $\Delta y$ & $N_y$ & $q_0$ & $\lambda_q$ & $S$ & $q_{\text{BG}}$ & $x_0$ & $f_X$ & Integral \\
                $(\text{m})$ & $(-)$ & $(\text{m})$ & $(-)$ & $(\text{W} \times \text{m}^{-2})$ & $(\text{m}$) & $(\text{m})$ & $(\text{W} \times \text{m}^{-2})$ & $(\text{m})$ & $(-)$ & $(\text{W} \times \text{m}^{-1})$
                \\\hline\hline
                \textcolor{blue}{2.3E$-$03} & \textcolor{blue}{15} & \textcolor{blue}{2.3E$-$03} & \textcolor{blue}{13} & 6.11E$+$06 &5.35E$-$03 &2.11E$-$03 &2.53E$+$05 &1.38E$-$03 &3.28E$+$00 & 1.59E$+$05 \\\hline
                \textcolor{blue}{1.1E$-$03} & \textcolor{blue}{30} & \textcolor{blue}{1.1E$-$03} & \textcolor{blue}{26} & 9.16E$+$06 &5.76E$-$03 &1.70E$-$03 &2.69E$+$05 &2.72E$-$03 &3.17E$+$00 & 2.23E$+$05 \\\hline
                \textcolor{blue}{5.5E$-$04} & \textcolor{blue}{60} & \textcolor{blue}{5.5E$-$04} & \textcolor{blue}{53} & 1.11E$+$07 &6.68E$-$03 &1.33E$-$03 &2.45E$+$05 &4.13E$-$03 &2.87E$+$00 & 2.64E$+$05 \\\hline
                \textcolor{blue}{2.7E$-$04} & \textcolor{blue}{120} & \textcolor{blue}{2.7E$-$04} & \textcolor{blue}{107} & 1.19E$+$07 &7.57E$-$03 &9.48E$-$04 &1.96E$+$05 &4.98E$-$03 &2.69E$+$00 & 2.83E$+$05 \\\hline
                \textbf{\textcolor{blue}{1.4E$-$04}} & \textbf{\textcolor{blue}{240}} & \textbf{\textcolor{blue}{1.4E$-$04}} & \textbf{\textcolor{blue}{207}} & \textbf{1.24E$+$07} & \textbf{1.04E$-$02} & \textbf{9.27E$-$04} & \textbf{3.50E$+$04} & \textbf{5.34E$-$03} & \textbf{2.10E$+$00} & \textbf{2.75E$+$05} \\\hline
                \textcolor{blue}{6.8E$-$05} & \textcolor{blue}{480} & \textcolor{blue}{6.8E$-$05} & \textcolor{blue}{426} & 1.26E$+$07 &1.03E$-$02 &7.81E$-$04 &2.29E$-$08 &5.52E$-$03 &2.15E$+$00 & 2.78E$+$05 \\\hline
            \end{tabular}
            }
            \label{tab:space_convergence_eich}

            \centering
            \caption{Eich fitting parameters \cite{Eich_2013} and Eich integral (Eq. (\ref{eq:eich_integral})) as a function of $\Delta y \neq \Delta x$ with fixed $\Delta x= 5.5 \times 10^{-4}$ m and $\Delta t = 7.2 \times 10^{-6}$ s.}
            \renewcommand*\arraystretch{1.4}
            \resizebox{\textwidth}{!}{
            \begin{tabular}{|c|c|c|c||c|c|c|c|c|c|c|}
                \hline
                $\Delta x$ & $N_x$ & $\Delta y$ & $N_y$ & $q_0$ & $\lambda_q$ & $S$ & $q_{\text{BG}}$ & $x_0$ & $f_X$ & Integral \\
                $(\text{m})$ & $(-)$ & $(\text{m})$ & $(-)$ & $(\text{W} \times \text{m}^{-2})$ & $(\text{m}$) & $(\text{m})$ & $(\text{W} \times \text{m}^{-2})$ & $(\text{m})$ & $(-)$ & $(\text{W} \times \text{m}^{-1})$
                \\\hline\hline
                5.5E$-$04 & 60 & \textcolor{blue}{2.0E$-$03} & \textcolor{blue}{14} & 6.37E$+$06 &5.84E$-$03 &1.99E$-$03 &2.34E$+$05 &1.76E$-$03 &3.07E$+$00 & 1.63E$+$05\\\hline
                5.5E$-$04 & 60 & \textcolor{blue}{1.0E$-$03} & \textcolor{blue}{29} & 9.17E$+$06 &6.08E$-$03 &1.67E$-$03 &2.53E$+$05 &3.01E$-$03 &3.03E$+$00 & 2.21E$+$05\\\hline
                5.5E$-$04 & 60 & \textcolor{blue}{5.0E$-$04} & \textcolor{blue}{58} & 1.08E$+$07 &6.45E$-$03 &1.20E$-$03 &2.34E$+$05 &4.28E$-$03 &2.99E$+$00 & 2.57E$+$05\\\hline
                5.5E$-$04 & 60 & \textcolor{blue}{2.5E$-$04} & \textcolor{blue}{116} & 1.15E$+$07 &8.21E$-$03 &9.77E$-$04 &1.84E$+$05 &5.05E$-$03 &2.49E$+$00 & 2.73E$+$05\\\hline
                5.5E$-$04 & 60 & \textcolor{blue}{1.3E$-$04} & \textcolor{blue}{232} & 1.19E$+$07 &1.01E$-$02 &8.73E$-$04 &1.78E$+$04 &5.37E$-$03 &2.17E$+$00 & 2.63E$+$05\\\hline
                5.5E$-$04 & 60 & \textcolor{blue}{6.3E$-$05} & \textcolor{blue}{464} & 1.21E$+$07 &1.02E$-$02 &7.65E$-$04 &2.57E$-$12 &5.53E$-$03 &2.16E$+$00 & 2.67E$+$05\\\hline
                5.5E$-$04 & 60 & \textcolor{blue}{3.1E$-$05} & \textcolor{blue}{928} & 1.22E$+$07 &1.03E$-$02 &7.21E$-$04 &4.47E$-$09 &5.60E$-$03 &2.16E$+$00 & 2.71E$+$05\\\hline
            \end{tabular}
            }
            \label{tab:space_convergence_eich_dy}
            
        \end{table}

    \subsection{Estimate of the accuracy of FAHF: comparison against COMSOL Multiphysics\textsuperscript{\textregistered}}\label{sec:comparison}

        Downstream proving the choice of $\Delta t = 7.2 \times 10^{-6}$ s and $\Delta x = \Delta y = 1.4 \times 10^{-4}$ m as acceptable in terms of convergence errors (Sec. \ref{sec:space_time_convergence}), energy balance error (Sec. \ref{sec:energy_balance}), and heat flux behaviour (Sec. \ref{sec:sensitivity_Eich}), the necessary condition for a well-posed cross-code comparison of this case against COMSOL is ensured.

        \subsubsection{Assessment of the geometrical error in FAHF: CH\#2 $-$ bulk of the tile}\label{sec:comparison_fahf_comsol2d_comsol3d}

\begin{figure}
                \centering
                
                \subfloat[]{\includegraphics[width = 0.485\textwidth]{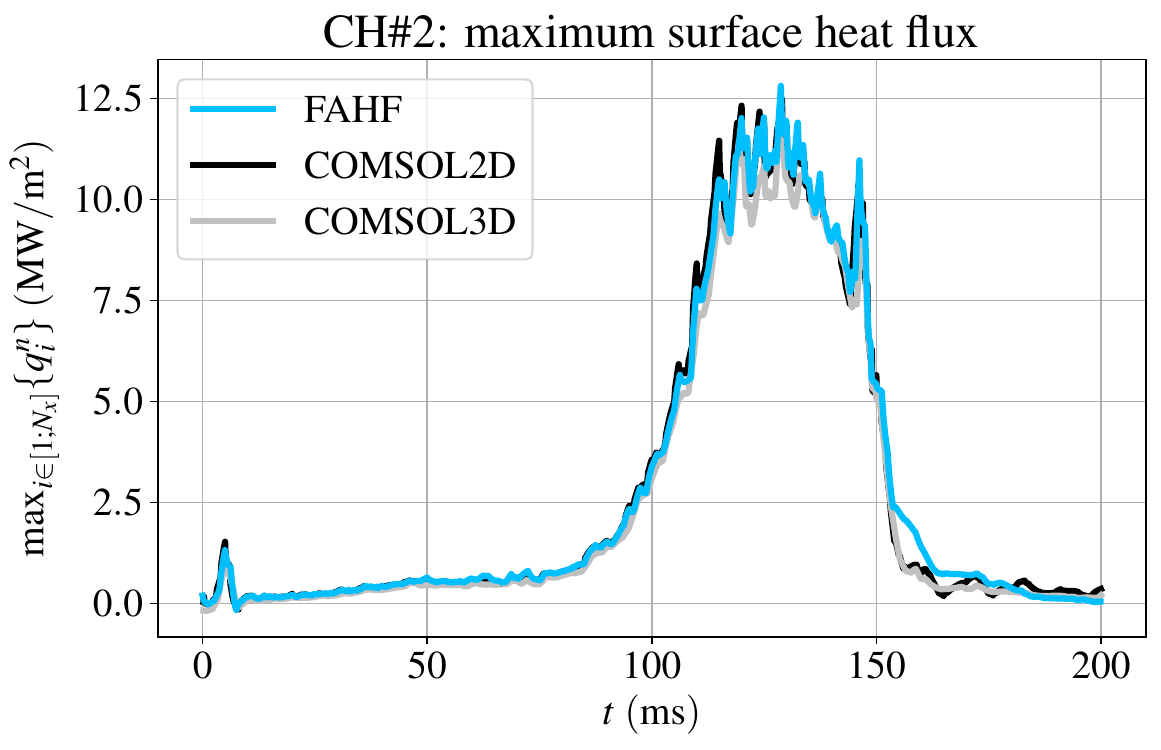}\label{fig:fahf_comsol2d_comsol3d-c}}
                \hspace{0.15 cm}
                \subfloat[]{\includegraphics[width = 0.485\textwidth]{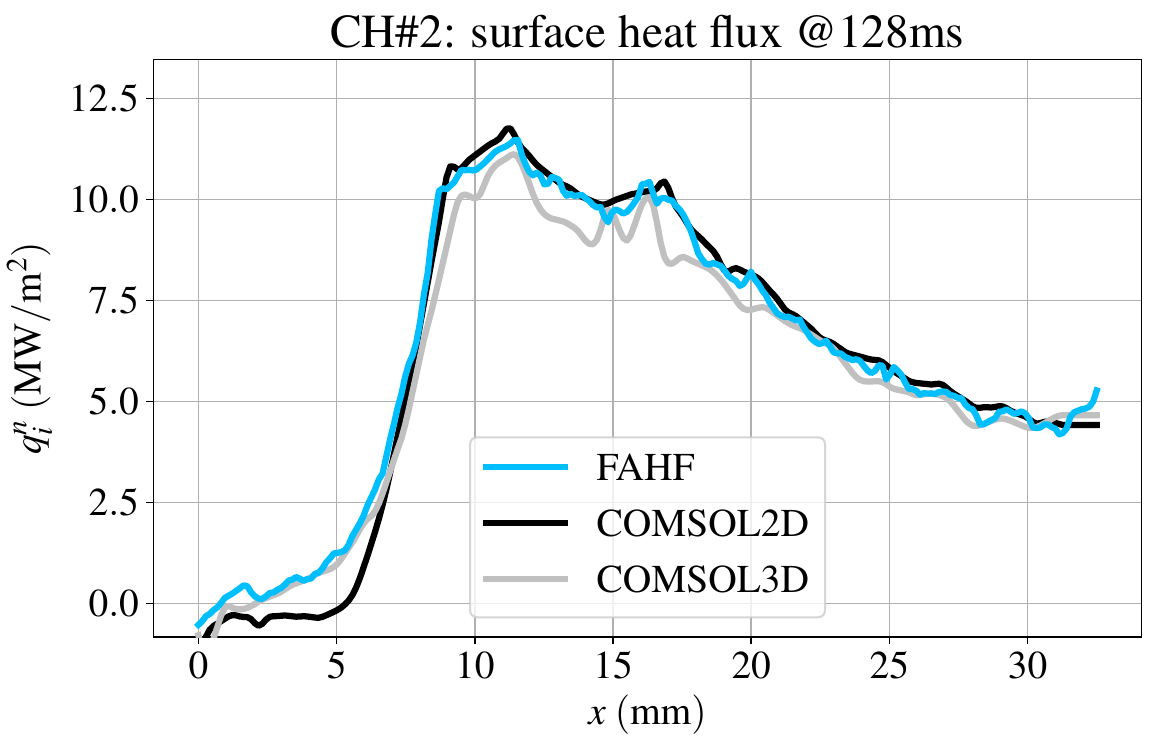}\label{fig:fahf_comsol2d_comsol3d-d}}\\
                \subfloat[]{\includegraphics[width = 0.485\textwidth]{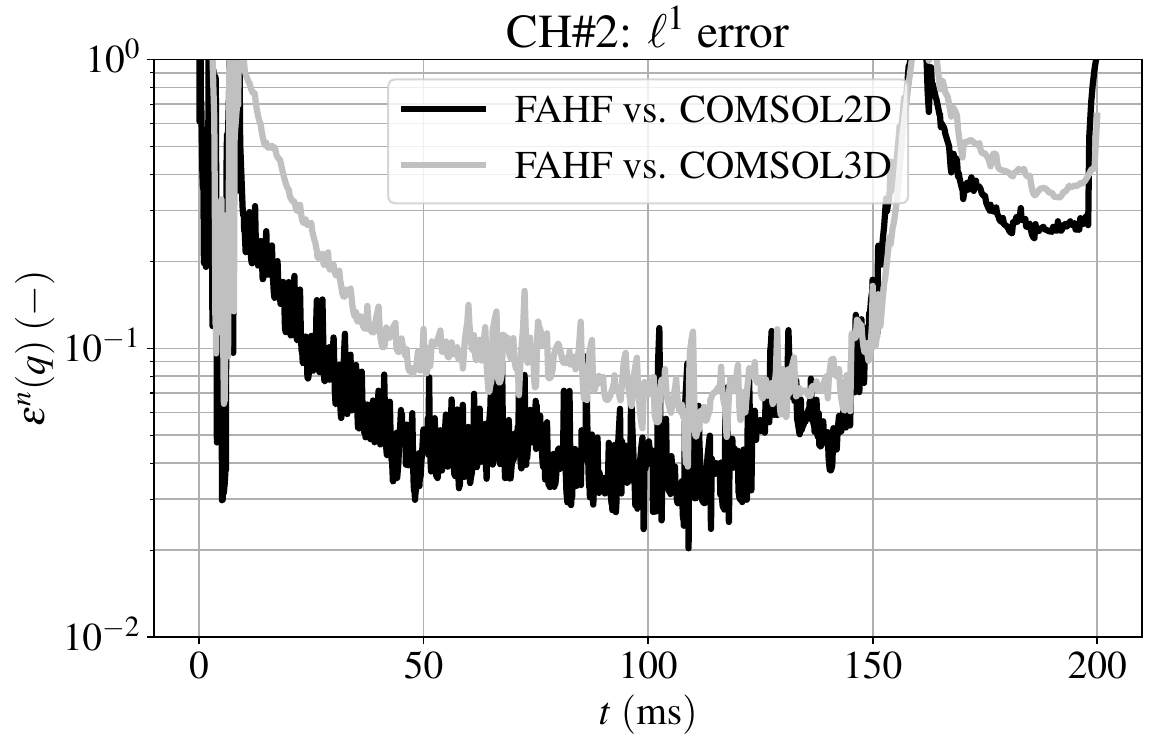}\label{fig:fahf_comsol2d_comsol3d-a}}
                \hspace{0.15 cm}
                \subfloat[]{\includegraphics[width = 0.485\textwidth]{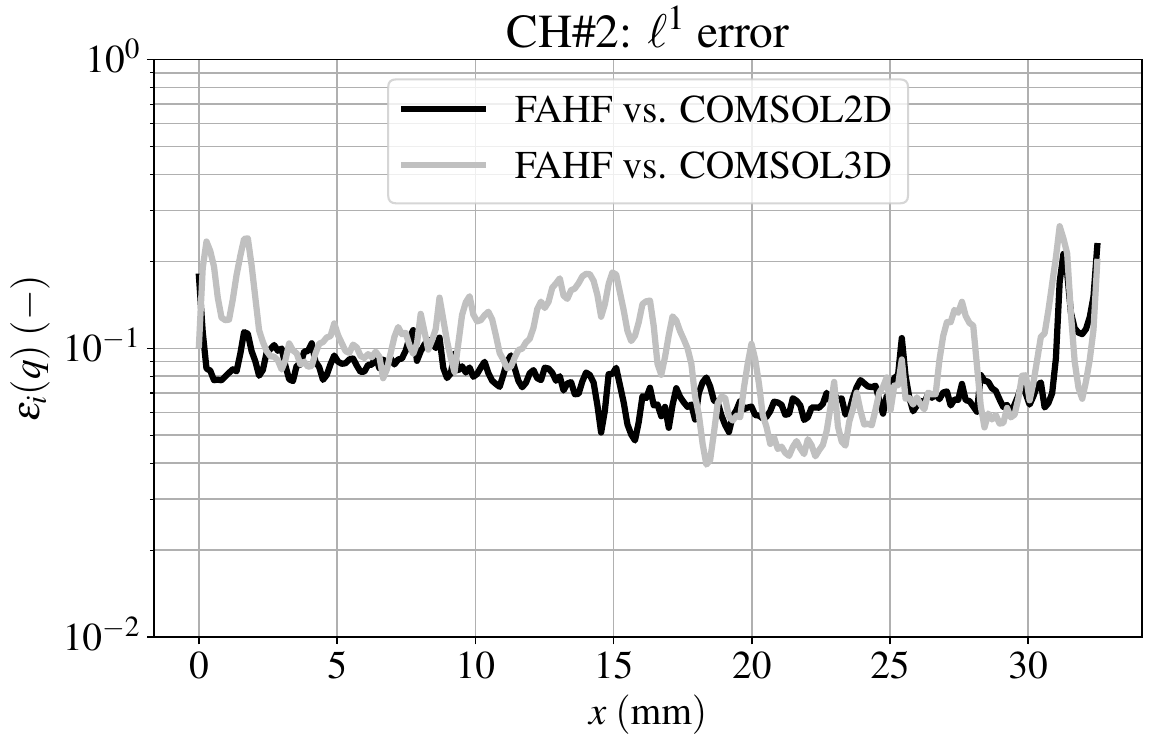}\label{fig:fahf_comsol2d_comsol3d-b}}\\
                
                \subfloat[]{\includegraphics[width = 0.485\textwidth]{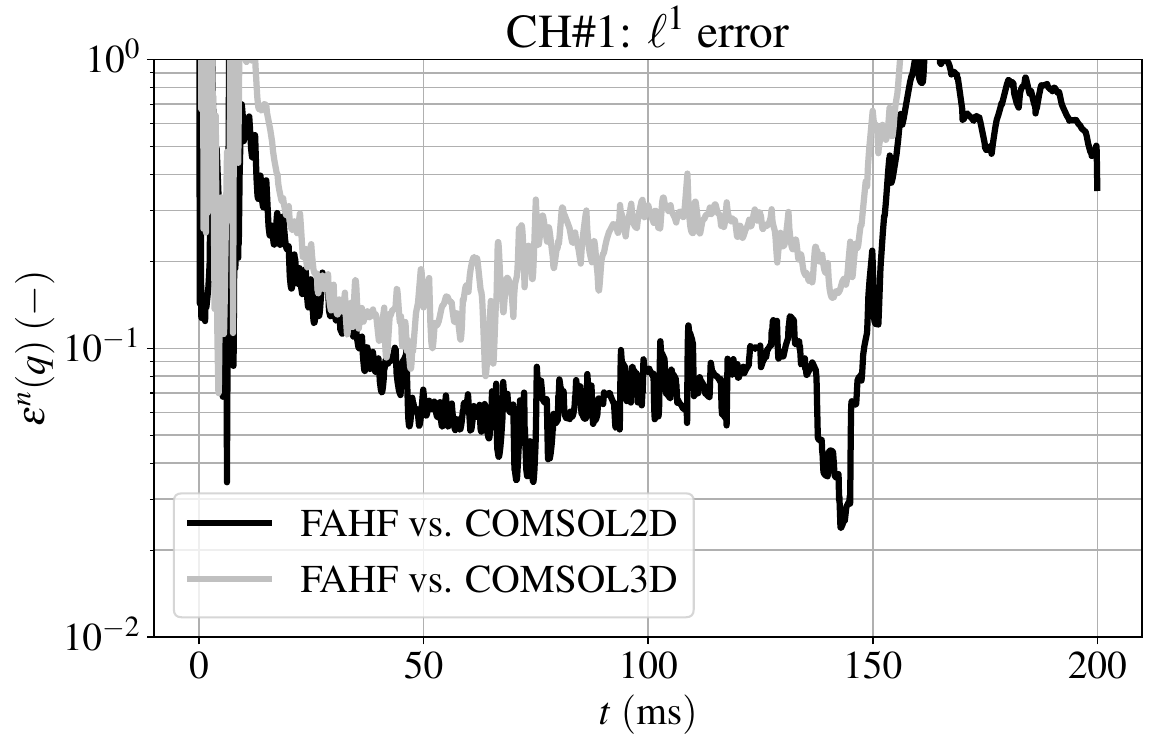}\label{fig:fahf_comsol2d_comsol3d_ch1-a}}
                \hspace{0.15 cm}
                \subfloat[]{\includegraphics[width = 0.485\textwidth]{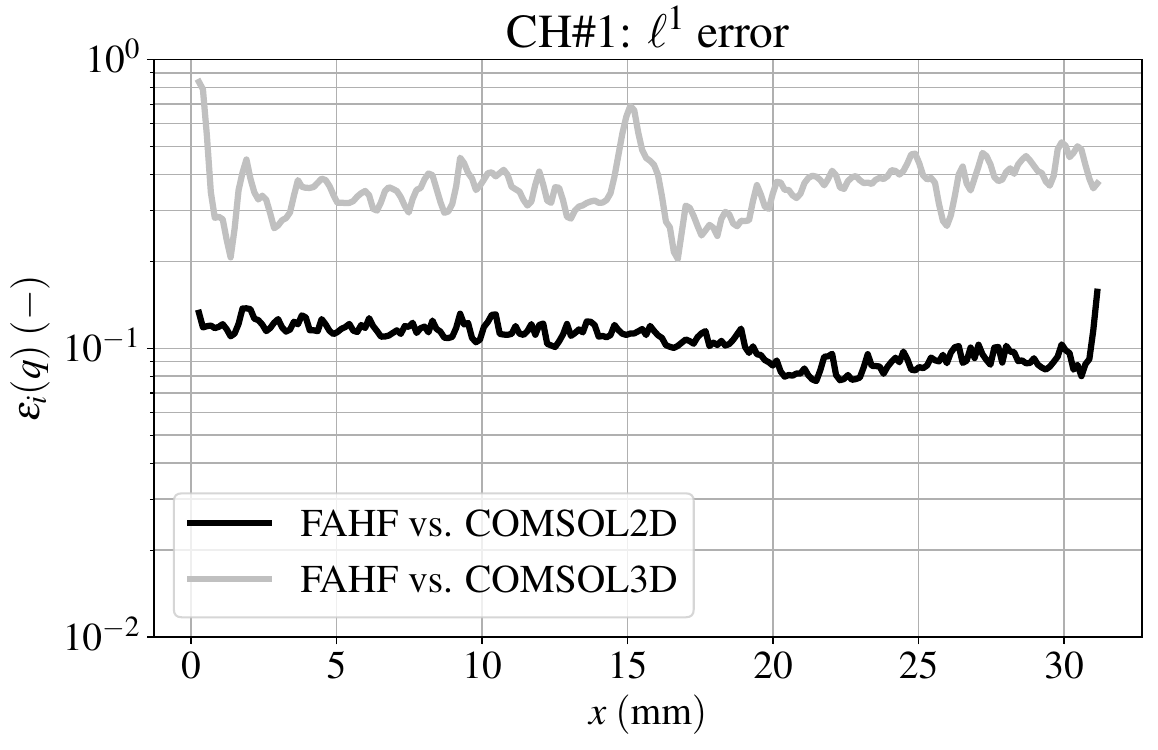}\label{fig:fahf_comsol2d_comsol3d_ch1-b}}

                \caption{CH\#2 $-$ bulk of the tile: (a) maximum heat flux in time $\text{max}_{i \in [1;N_x]}\{q_i^n\}$; (b) heat flux profiles in space at time instant $t_* = 128$ ms over the reference sub-chord; (c) error $\varepsilon^n(q)$ in time from Eq. (\ref{eq:l1_norm_time}); (d) error $\varepsilon_i(q)$ in space from Eq. (\ref{eq:l1_norm_space}). CH\#1 $-$ vicinity of poloidal-running gap: (e) error $\varepsilon^n(q)$ in time from Eq. (\ref{eq:l1_norm_time}); (f) error $\varepsilon_i(q)$ in space from Eq. (\ref{eq:l1_norm_space}).}  
                
                \label{fig:fahf_comsol2d_comsol3d}
            \end{figure}

           According to Sec. \ref{sec:comsol_setup}, COMSOL is run in two dimensions (COMSOL2D) and three dimensions (COMSOL3D), with temperature-independent material properties and with a single layer of 29-mm molybdenum, hence matching FAHF's Asms. \#1 and \#2. This allows insights in the geometrical error alone, as implied by Asm. \#3. In the present section, simulation data are computed on CH\#2 of Sec. \ref{sec:ref_chord}, while Sec. \ref{sec:comparison_fahf_comsol2d_comsol3d_ch1} deals with the CH\#1 instance.
           \\The satisfactory agreement between FAHF, COMSOL2D and COMSOL3D on CH\#2 is pictured in Fig. \ref{fig:fahf_comsol2d_comsol3d}, which shows the comparison of: (a) the maximum heat flux in time, $\text{max}_{i \in [1;N_x]}\{q_i^n\}$; (b) the heat flux profiles in space at time instant $t_* = 128$ ms; (c) the error $\varepsilon^n(q)$ in time from Eq. 
           (\ref{eq:l1_norm_time}); (d) the error $\varepsilon_i(q)$ in space from Eq. (\ref{eq:l1_norm_space}). In particular, (a) and (b) allow for useful qualitative insights, aided by the lack of camera saturation at $t_* = 128$ ms along CH\#2. 
           \\According to Fig. \ref{fig:fahf_comsol2d_comsol3d-a}, the agreement between FAHF and COMSOL2D holds for the entire time window $[40;140]\;\text{ms}$, with an average error of $\varepsilon^n(q) = 5\%$. For FAHF vs. COMSOL3D, $\varepsilon^n(q)$ is clearly higher than its 2D counterpart, though still being an acceptable $8\%$.
           \\A significant inaccuracy of FAHF is observed when the plasma: is started ($t \sim 10$ ms); reverts back to being limited on the centre column ($t \sim 160$ ms, hence suddenly depositing its power elsewhere than the divertor, Sec. \ref{sec:st40}); is terminated ($t \sim 200$ ms), i.e. when evolving on a short timescale\footnote{The maximum heat flux fluctuations in $[100;150]$ ms are a combination between: saturation of $T_{\text{IR}}$ (Sec. \ref{sec:ref_shot}); considering the only maximum heat flux; plasma vertical oscillation. These do not fall under the umbrella of 'fast dynamics'.} (Fig. \ref{fig:fahf_comsol2d_comsol3d-c}). Accurately capturing fast transients would indeed require proper numerical handling \cite{Adebayo-Ige_2024}, but it is not of concern in the present study, where the $\Delta t_{\text{IR}}$ (Table \ref{tab:camera}) would not be suited in the first place.
           \\Fig. \ref{fig:fahf_comsol2d_comsol3d-b} suggests that similar conclusions hold for the error in space $\varepsilon_i(q)$, which satisfactorily attains values of $8\%$ for FAHF vs. COMSOL2D all throughout the bulk of the tile, i.e. $x \in [2.5;30]$ mm. The enhanced error at the boundaries is imputable to FAHF there reverting to first-order BCs (Sec. \ref{sec:BCs}). Slightly higher values of $10\%$, though still acceptable, are found with respect to COMSOL3D.
           \\Given the supportive evidence above, the appropriateness of adopting a 2D approach in FAHF (Asm. \#3), rather than a full 3D treatment, is confirmed for chords in the bulk of the tile, coherently with the findings of Ref.\cite{marsden2024}. This conclusion especially holds during the time window of most interest, and its robustness is further enhanced by considering the implicit over-estimation of the error according to Sec. \ref{sec:heat_flux_error}.
            
        \subsubsection{Assessment of the geometrical error in FAHF: CH\#1 $-$ vicinity of poloidal-running gap}\label{sec:comparison_fahf_comsol2d_comsol3d_ch1}

            Conversely to Sec. \ref{sec:comparison_fahf_comsol2d_comsol3d}, results on CH\#1 are used in Fig. \ref{fig:fahf_comsol2d_comsol3d_ch1-a} and \ref{fig:fahf_comsol2d_comsol3d_ch1-b}. Only the errors are shown, for the sake of conciseness.
            \\Average $\varepsilon^n(q)$ values of $7\%$ for FAHF vs. COMSOL2D and $22\%$ vs. COMSOL3D are found in $[40;140]$ ms on CH\#1, hence above their counterparts on CH\#2. The same holds for $\varepsilon_i(q)$ in space, with $11\%$ for the FAHF vs. COMSOL2D instance, and $36\%$ vs. COMSOL3D, above the values on CH\#2.
            \\This behaviour follows the expectations, for any poloidal-running gaps violates the assumption of axisymmetry along the toroidal direction leveraged by Asm. \#3. The 2D approach is therefore shown to be more appropriate for chords away from the poloidal-running gaps ($\sim$ CH\#2) than for chords in their proximity ($\sim$ CH\#1).
            \\This result should also caution on the selection of a representative region of interest, were a single-chord approach adopted (Sec. \ref{sec:dimensionality}).
            
        \subsubsection{Assessment of the model errors in FAHF}\label{sec:comparison_fahf_comsol2d}

\begin{figure}
                \centering
                \subfloat[]{\includegraphics[width = 0.485\textwidth]{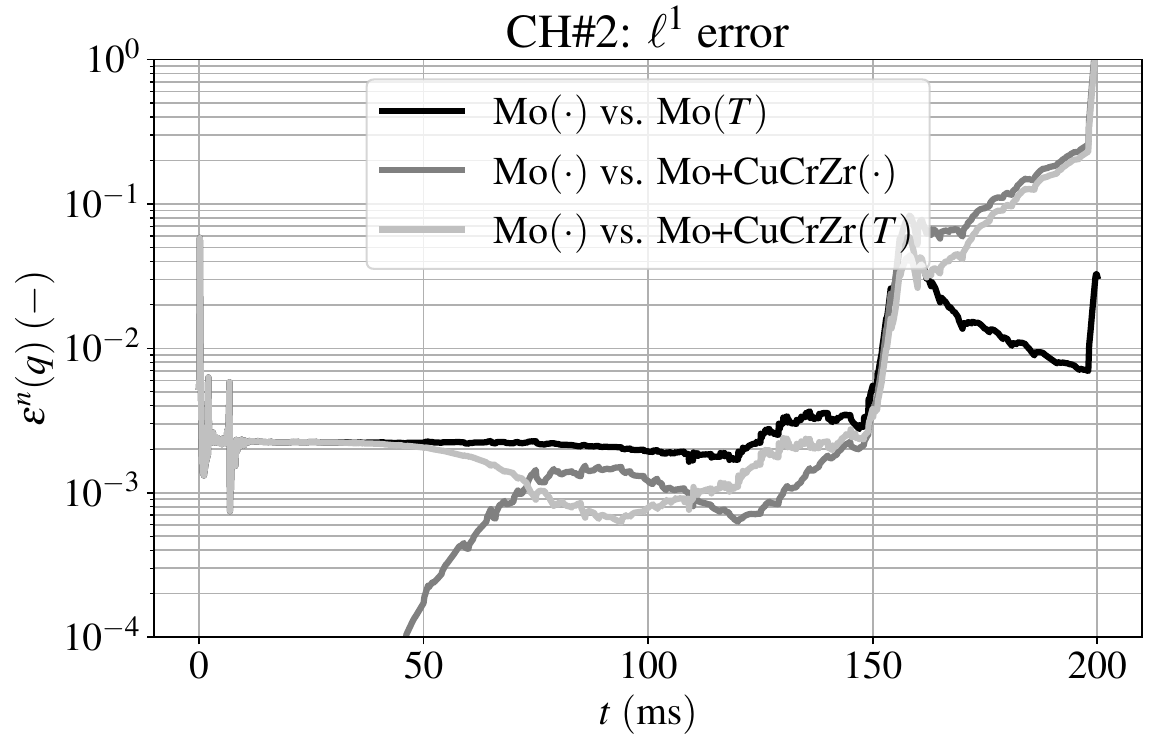}\label{fig:fahf_comsol2d-a}}
                \hspace{0.15 cm}
                \subfloat[]{\includegraphics[width = 0.485\textwidth]{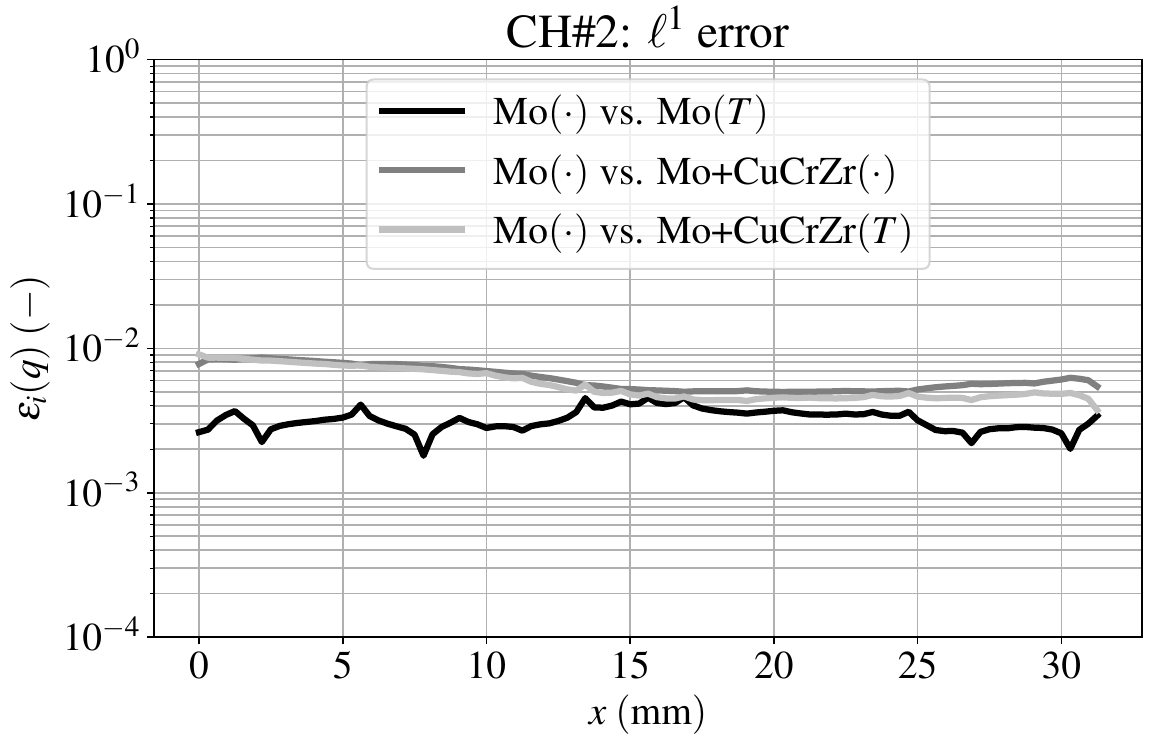}\label{fig:fahf_comsol2d-b}}
                \caption{(a) Error $\varepsilon^n(q)$ in time from from Eq. (\ref{eq:l1_norm_time}) and (b) error $\varepsilon_i(q)$ in space from Eq. (\ref{eq:l1_norm_space}), computed among the two-dimensional COMSOL cases with respect to the COMSOL simulation implementing FAHF-like models.}  
                \label{fig:fahf_comsol2d}
            \end{figure}

            Starting from the COMSOL2D simulation on CH\#2 of Sec. \ref{sec:comparison_fahf_comsol2d_comsol3d} with (i) temperature-independent material properties and a single layer of 29-mm molybdenum (Mo$(\cdot)$), the set of simulations is enlarged to assess the model simplifications in Asms. \#1 and \#2. COMSOL is indeed run in 2D on CH\#2: (ii) with temperature-dependent material properties \cite{mo_rho, mo_cp, mo_k} and a single layer of 29-mm molybdenum (Mo$(T)$); (iii) with temperature-independent material properties and a double layer of 4-mm molybdenum and 25-mm copper-cromium-zirconium (Mo+CuCrZr$(\cdot)$); (iv) with temperature-dependent material properties \cite{cucrzr_rho_cp_k} and a double layer of 4-mm molybdenum and 25-mm copper-cromium-zirconium (Mo+CuCrZr$(T)$).
            \\FAHF, already compared against COMSOL2D in Sec. \ref{sec:comparison_fahf_comsol2d_comsol3d}, is excluded from this comparison, to allow for the precise appreciation of the minute changes here involved, which would be obscured otherwise. The errors according to Eq. (\ref{eq:l1_norm_time}) and (\ref{eq:l1_norm_space}) are therefore computed with the COMSOL result Mo($\cdot$) taken as the reference $\tilde{q}$.
            \\In particular, $\varepsilon^n(q) < 0.5 \% \; \forall t \in [15;150]$ ms in Fig. \ref{fig:fahf_comsol2d-a}, and $\varepsilon_i(q) < 1\% \; \forall x$ in Fig. \ref{fig:fahf_comsol2d-b}. This confirms that Asms. \#1 and \#2 are appropriate in the present case. Implementing more sophisticated models in FAHF could be accomplished by leveraging its modularity, were this be needed.
            \\Importantly, $\varepsilon^n(q)$ is practically zero if $t \lesssim 50$ ms for the case Mo($\cdot$) vs. Mo+CuCrZr($\cdot$). This confirms that any thermal information indeed takes tens of ms to reach the Mo-CuCrZr interface, located 4 mm in the depth of the divertor tiles, hence consistently with Sec. \ref{sec:theory}.
            \\Also noticed in Sec. \ref{sec:comparison_fahf_comsol2d_comsol3d}, the peculiar time windows of $t \lesssim 10$ ms and $t \gtrsim 160$ ms (of secondary importance in these analyses) violate the conclusions above, and are further commented in Sec. \ref{sec:geometrical_error_time}.
            \\Notably, the computational time of the COMSOL simulation Mo$(\cdot)$, with settings of Sec. \ref{sec:comsol_setup}, is a factor 8 bigger than FAHF's, notwithstanding the pre-processing time during the data handling, easily optimised in FAHF via Python tools \cite{python}.

%%%%%%%%%%%%%%%%%%%%%%%%%%%%%%%%%%%%%%%%%%%%%%%%%%%%%%%%%%
%%%%%%%%%%%%%%%%%%%%%%%%%%%%%%%%%%%%%%%%%%%%%%%%%%%%%%%%%%

\section{Discussion}\label{sec:discussion}

        \subsection{Convergence properties of the energy balance error}\label{sec:energy_balance_convergence}
        
            The goal of this paragraph is to discuss the behaviour of $\varepsilon_*(\mathcal{E})$ of Sec. \ref{sec:energy_balance}, with $\Delta x = \Delta y$. The underlying algebra is explicitly detailed in Appendix \ref{sec:algebra_energy}, and states how the error $\varepsilon_*(\mathcal{E})$ must abide to a first order convergence in space, the cases of Table \ref{tab:space_convergence} being considered.
            \\On the contrary, an explicit dependence on $\Delta t$ is almost totally absent in Table \ref{tab:time_convergence} (last column) suggesting that $o(\Delta t) \lesssim o(\Delta y)$, i.e. the trend in time of $\varepsilon_*(\mathcal{E})$ is hampered by $o(\Delta y)$.
            \\The first-order error $o(\Delta y)$ affecting the heat flux $q_i^n$ via Eq. (\ref{eq:q_1st}) is indeed found to be the dominant contribution, with regards to energy conservation. Although the resulting energy error is still acceptable (Sec. \ref{sec:energy_balance}), a point of improvement for FAHF would be using higher-order discretisations in the fashion of Refs.\cite{theodor, ASDEX_IR_2}'s instead of Eq. (\ref{eq:q_1st}), were higher precision in energy required.
            
        \subsection{Implications on choosing $\Delta x$ and $\Delta y$}\label{sec:implications_dx_dy}

            As also suggested by Ref.\cite{Adebayo-Ige_2024}, Sec. \ref{sec:sensitivity_Eich_dy} clarifies how a precise heat flux can be calculated at a lower cost with a proper choice of $\Delta y < \Delta x$, without the need of an extremely small $\Delta x$.
            \\However, the assessment of the error in $q_i^n$ and in the energy balance (Sec. \ref{sec:energy_balance}) as a function of $\Delta x$ is also recommended. For small enough $\Delta y$, $\Delta x$ becomes the bottleneck in $o(\Delta x^2) + o(\Delta y)$ of Eqs. (\ref{eq:q_1st}) and (\ref{eq:error_energy_time}), potentially even more so if higher-order discretisations of $q_i^n$ with $\Delta y$ are employed.
            \\From the considerations above, the ultimate $\Delta x$ selected might fall below the spatial resolution of the camera, i.e. $\Delta x < \Delta x_{\text{IR}}$. This is not of concern as long as a linear interpolation of $T_{\text{IR}}$ is employed (Sec. \ref{sec:BCs}). If, on the one hand, $\Delta x < \Delta x_{\text{IR}}$ does not add any physics-relevant information, on the other hand, it plays a role in the numerics of the problem. And the cautious user should not be restrained by $\Delta x_{\text{IR}}$ when identifying the most proper $\Delta x$.

        \subsection{Implications on choosing $\Delta x_{\text{IR}}$}\label{sec:implications_dxIR}
            
            The fact that the resolution along $y$ (independent on the diagnostic apparatus) mostly determines the precision of $q_i^n$ does loosen the constraints on the required IR camera resolution $\Delta x_{\text{IR}}$ (Sec. \ref{sec:ir_camera_hardware}).
            \\While $\Delta x$ follows from Sec. \ref{sec:implications_dx_dy}, the suitable $\Delta x_{\text{IR}}$ is such that to allow capturing features of $T_{\text{IR}}$, a smoother and broader profile than the corresponding surface heat flux.
            \\Nevertheless, power fall-off lengths significantly narrower than the Eich scaling predictions \cite{zhang2024experimental, COMPASS_lq, TCV_lq} do rise the concern on the resolution $\Delta x_{\text{IR}}$ required by IR thermographic systems in current and future devices. This clearly reverberates on the choice of $\Delta x$ and $\Delta y$ (Sec. \ref{sec:implications_dx_dy}).
            \\Last, the FAHF result $q_i^n$ is seen to very closely follow $q_{\text{Eich}} \; \forall \Delta x \geq 5.5 \times 10^{-4}$ m in Fig. \ref{fig:q_space-a}. Instead, a deviation from the Eich fit in $x \in [10;20]$ mm is noticeable $\forall \Delta x \leq 2.7 \times 10^{-4}$ m, and also predicted by COMSOL in Fig. \ref{fig:fahf_comsol2d_comsol3d-d}. This deviation is a genuine feature because $\Delta x > \Delta x _{\text{IR}} = 2.1 \times 10^{-4}$ m (Sec. \ref{sec:ir_camera_hardware}). The effective spatial resolution of the ST40 IR system does allow for the appreciation of fine-scale structures in the heat flux footprint, not dissimilar for those predicted by the XGC code\cite{Matteo_PSI, Chang_2017}, meritorious of further studies.

        \subsection{Geometrical error}\label{sec:geometrical_error}

            Although Sec. \ref{sec:comparison_fahf_comsol2d_comsol3d} already concluded on the appropriateness of Asm. \#3 in FAHF, further investigations are warranted in the light of Sec. \ref{sec:comsol_setup}. Only CH\#2 of Sec. \ref{sec:ref_chord}, most suitable for a 2D analysis (Sec. \ref{sec:comparison_fahf_comsol2d_comsol3d}).
        
            \subsubsection{Features in time}\label{sec:geometrical_error_time}
               
               The error FAHF vs. COMSOL3D of $10\%$ in $[40;90]$ ms is greater than the $7\%$ in $[90;140]$ ms. The same is not true for the error FAHF vs. COMSOL2D instead, which reads $5\%$ in both $[40;90]$ ms and $[90;140]$ ms. This 3D-related feature likely follows from the appearance of the strike point in view of the camera during $[90;140]$ ms, which leads to $|\partial_xT_{\text{IR}}| \gg |\partial_{\phi}T_{\text{IR}}|$ (Sec. \ref{sec:dimensionality}), i.e. a stronger tendency for the heat flux to flow poloidally (favouring a 2D treatment, hence lowering the error) rather than toroidally (inviting 3D flows, thus enhancing the error).
               \\Additionally, COMSOL3D maximum is systematically in defect of its COMSOL2D counterpart all throughout $[40;140]\;\text{ms}$. COMSOL3D hence predicts that the same PFC temperature (i.e. $T_{\text{IR}}$) tends to be reached by means of a lower surface heat flux along $y$. This effect is again tied to the three-dimensionality of the problem, and presumably due to the additional heat flux that flows along $\phi$ (i.e. across the chords, Sec. \ref{sec:dimensionality}) towards the CH\#2 in COMSOL3D (and in reality), by virtue of $\partial_{\phi}T_{\text{IR}} < 0$ (Fig. \ref{fig:bundle-b}). This additional contribution along $\phi$, accounted neither by FAHF nor by COMSOL2D, helps rising the temperature along CH\#2 in the 3D case, thus requiring a lower $q_i^n$ along $y$.
               \\Moreover, because of the FAHF incapability of resolving rapid dynamics, FAHF errors diverge towards 100\% during the fast transients at the instants $t \sim 10$ ms and $t \sim 160$ ms, over a time of $\sim 5$ ms. This error is a thermal information which, in the legitimate approximation of semi-infinite solid \cite{incropera} (Sec. \ref{sec:theory}), affects a region of thickness $2 \sqrt{\alpha \times 5 \; \text{ms}} \sim 1$ mm along $y$. According to the resistance-capacitor circuital interpretation\cite{incropera}, this information then exponentially dissipates with a time constant in the order of $(1 \; \text{mm})^2 / \alpha \sim 19$ ms. And the FAHF error is indeed seen to drop over a timescale of this magnitude, following the $t \sim 10$ ms and $t \sim 160$ ms events. The exponential fit of the neat FAHF vs. COMSOL3D error over the time window $[10;35]$ ms returns a time constant of 12 ms with a satisfactory $R^2 = 98 \%$. This crucially implies that the error during the initial phase ($t \sim 10$ ms) does not detrimentally impact the longer-term dynamics predicted by FAHF (loss of memory).
               \\On the contrary, the error at $t \gtrsim 160$ ms, despite firstly dropping, remains significant. This is presumably implied by the nature of this phase, an inertial cooling after the plasma magnetic topology reverts back to limited (with a sudden drop of divertor heating). Such a simple dynamics of inertial cooling, especially in conjunction with adiabatic boundaries (Sec. \ref{sec:heat_eq_theory}), is highly sensitive to the conditions at its beginning ($t \sim 160$ ms), which are very different between FAHF and COMSOL ($\varepsilon^n(q) \sim 100$\%).

           \subsubsection{Features in space}\label{sec:geometrical_error_space}
               
               The FAHF vs. COMSOL3D error in space $\varepsilon_i(q)$ increases (i) towards the toroidal-running gaps, i.e. $x = 0$ and $x = 32$ mm. This is expected because of the first-order boundary conditions of FAHF (Sec. \ref{sec:BCs}), and because of the higher degree of three-dimensionality necessarily involved around these regions: the adiabatic boundary invites the heat to flow elsewhere than along $x$, e.g. along $\phi$, not modelled in 2D. Because further physical effects thereabouts occurring (e.g. tile side-heating due to finite ion Larmor radius \cite{marsden2024, GUNN201775}), however not included in any thermographic inversion, (i) must be under-emphasised anyway, and is therefore not of concern.
               \\Nonetheless, $\varepsilon_i(q)$ is also enhanced (ii) around the centre of the tile, i.e. $x \sim 15$ mm. This can be ascribed to the dual of (i): the centre of the tile allows for a 3D heat flux behaviour by offering the loosest constraints, by virtue of being the point farthest away from any adiabatic boundary. A gradual transition towards 2D between these 3D extremes (i) and (ii) instead characterises the remaining regions, where the error $\varepsilon_i(q)$ drops.

           \subsubsection{Negative heat fluxes}\label{sec:negative_q}
           
               The detection of negative heat fluxes in IR thermographic inversions has been historically ascribed to (or a combination of) mechanical vibrations \cite{CMOD_IR, KSTAR_IR_2}, impurity surface layers \cite{Adebayo-Ige_2024, hermann} and black-body radiation emission from the PFCs \cite{hermann}, at least.
               \\Negative surface heat fluxes are here predicted in all the three cases: (i) in $(0;10]$ ms in Fig. \ref{fig:fahf_comsol2d_comsol3d-c}; (ii) at $t_* = 128$ ms on the left of the peak in Fig. \ref{fig:fahf_comsol2d_comsol3d-d}, $x \in [0;5]$ mm; (iii) all throughout the the divertor cooling phase (Sec. \ref{sec:geometrical_error_time}) at $t \gtrsim 160$ ms, not shown.
               \\With regards to (i), vibrations during the plasma start-up are the most reasonable explanation. The same does not instead hold for (ii) nor (iii), presumably affected by the presence of an impurity surface layer and non-negligible black-body emission. Both the above-mentioned points are worthwhile further investigating in the future, despite not impacting the conclusions of the present study.

%%%%%%%%%%%%%%%%%%%%%%%%%%%%%%%%%%%%%%%%%%%%%%%%%%%%%%%%%%
%%%%%%%%%%%%%%%%%%%%%%%%%%%%%%%%%%%%%%%%%%%%%%%%%%%%%%%%%%

\section{Conclusions and outlook}\label{sec:conclusions}

    Data of divertor surface temperature are recorded by ST40's IR camera system with the highest effective spatial resolution available worldwide ($0.21 \pm 0.03$ mm/pixel). Leveraging such highly-resolved data, FAHF, the new Tokamak Energy's tool for inverse thermographic analyses, is developed as a finite-difference explicit solver of the unsteady heat conduction equation. Under the simplifying assumptions of temperature-independent material properties, all-molybdenum mono-layered PFCs, and two-dimensionality, FAHF calculates the surface heat flux impinging on the divertor.
    \\In this work, we have demonstrated the suitability of FAHF as a tool for inverse thermographic analyses.
    \\The formal verification of the code by means of time and space convergence is successful, in fact showing the expected first- and second-order convergence with $\Delta t$ and $\Delta x = \Delta y$ (for the sake of simplicity), respectively. The most appropriate combination of $\Delta t$ and $\Delta x$ is chosen accordingly, such as to guarantee a satisfactory compromise between precision of the temperature output and computational cost.
    \\The verification is here further corroborated by an assessment of the energy balance. The error affecting the energy balance is evidenced to attain satisfactorily low values and: to feature a well-defined trend with $\Delta x$; to lack a dependence on $\Delta t$. Both such characteristics are successfully motivated in terms of ordering of the different contributions to the discretisation error, with $o(\Delta y)$ always being the dominant contribution. The energy-preserving properties of FAHF, whose assessment is recommended in the other codes, are hence deemed acceptable from the viewpoint of the internal consistency of the code, and for applications to power balance studies in tokamaks.
    \\The behaviour of the surface heat flux (along the divertor depth $y$) is also thoroughly scrutinised. The Eich fit is used to extract physical parameters of interest in the two separate cases of $\Delta x = \Delta y$ and $\Delta x \neq \Delta y$. On the one hand, in both instances a strong sensitivity of the fitting parameters as a function of the spatial resolution is found, with variations up to a factor 2.5. On the other hand, the two instances ultimately show a convergence of the fitting parameters towards common values, as the spatial resolution is enhanced. Converged values of the $\Delta x \neq \Delta y$ case are achieved at a lower computational cost, by virtue of the leading dependence of the surface heat flux being $\Delta y$ (divertor depth), rather than $\Delta x$ (along the divertor). Although this dependence is commonly leveraged in inverse thermographic analyses, the error implied by $\Delta x$ may become noticeable in the limit $\Delta y \ll \Delta x$, and detrimentally affect the surface heat flux. Routinely analysing the surface heat flux as a function of the space step $\Delta x$ along the divertor is thus recommended. This has consequences in dictating the appropriate spatial resolution of the IR thermographic system, which this study comments.
    \\With the above guaranteeing a satisfactory precision of FAHF, the comparison with the renowned COMSOL Multiphysics\textsuperscript{\textregistered} is well-posed, and evinces a satisfactory accuracy of FAHF. The average agreement within $8\%$ in time and $10\%$ in space among FAHF and COMSOL (3D) demonstrates the legitimacy of the 2D treatment adopted in FAHF. However, the agreement in space particularly worsens in the proximity of the poloidal-running gaps, and the resulting 22\% should discourage the application of a 2D treatment in such areas. The simplistic implementation in FAHF of temperature-independent material properties and single molybdenum layer is proven to be appropriate in ST40 shots: the corrections implied by including these features in COMSOL (2D) are shown to be of minor entity ($< 1\%$).
    \\Regions of inaccuracy in time of FAHF are circumscribed to the beginning/end of the plasma shot, and at the termination of the divertor heating phase. These fast dynamics could be properly resolved by improving the temporal discretisation properties of FAHF, but are outside the scope of the present work, outside the time window of interest for thermographic studies, and outside the time resolution of the system.
    \\In space, the error $\sim 20\%$ of FAHF is instead relegated to the regions in proximity of the toroidal-running gaps between divertor tiles. This is true in the 2D case, presumably because of the low-precision Neumann boundary conditions implemented in FAHF, and in the 3D case, also because of the adiabatic nature of these boundaries which invites a toroidal flow of the heat flux. Either instance does not pose a threat to the heat flux calculated by FAHF, as the above-mentioned areas are to be under-emphasised anyway.
    \\In conclusion, Tokamak Energy's FAHF does qualify as a trustworthy tool, internally consistent and capable of providing a satisfactory precision and accuracy at an affordable computational cost, modulo a proper choice of the inputs which this work identifies: $\Delta t = 7.2 \times 10^{-6}$ s and $\Delta x = \Delta y = 1.4 \times 10^{-4}$ m.
    \\In perspective, further studies are planned in ST40 with the two-fold goal of improving the IR thermographic inversion experimentally and numerically. The current IR camera saturation temperature has already been extended from $118$ \textcelsius{} to $200$ \textcelsius{}, and a detailed cross-comparison with Langmuir probe measurements is underway. The installation of a second IR camera is being considered, for the exploitation of a dual-band adaptor \cite{ir_dual} would allow ameliorating any dependence on the uncertain surface emissivity. The commissioning of the dedicated heat-flux testing facility in Tokamak Energy is ongoing, and will pave the way for a validation of FAHF in a controlled environment. Improvements to the numerics of FAHF to handle fast dynamics and impurity surface layers \cite{Adebayo-Ige_2024, theodor} are also taken into account, alongside the implementation of temperature-dependent material properties, and multi-layered PFCs.
    
%%%%%%%%%%%%%%%%%%%%%%%%%%%%%%%%%%%%%%%%%%%%%%%%%%%%%%%%%%
%%%%%%%%%%%%%%%%%%%%%%%%%%%%%%%%%%%%%%%%%%%%%%%%%%%%%%%%%%

\section*{Acknowledgments}
    The submitted manuscript has been co-authored by a contractor of the U.S. Government under contract DE-AC05-00OR22725. Accordingly, the U.S. Government retains a nonexclusive, royalty-free license to publish or reproduce the published form of this contribution, or allow others to do so, for U.S. Government purposes.
    \\We then thank OpenAI's language model, ChatGPT, for providing insightful editorial feedback on early drafts of this manuscript.
    \\Last, acknowledgment to the ever-precious help of David and Gabriela Hoffman, Sam Suchal, Teresa Thornton, Rosanne Monteiro and Anne Marie Frohock is also duely given.

\section*{Author declarations}

The authors have no conflicts to disclose. The authors' contribution statement using CRediT reads as follows. \textbf{Matteo Moscheni}: conceptualization (equal); methodology (equal); software (lead); formal analysis (lead); data curation (equal); visualization (lead); writing - original draft (lead); writing - review and editing (equal). \textbf{Erik Maartensson}: conceptualization (equal);  methodology (equal); software (equal); writing - review and editing (equal). \textbf{Matthew Robinson}: software (equal); formal analysis (equal). \textbf{Chris Marsden}: software (equal). \textbf{Adrian Rengle}: resources (lead); data curation (equal). \textbf{Andrea Scarabosio}: conceptualization (equal); supervision (equal); writing - review and editing (equal). \textbf{Patrick Bunting}: conceptualization (equal); resources (equal); writing - review and editing (equal). \textbf{Travis Kelly Gray}: conceptualization (equal); resources (equal); supervision (equal); writing - review and editing (equal). \textbf{Elena Vekshina}: supervision (equal); writing - review and editing (equal). \textbf{Xin Zhang}: supervision (equal). \textbf{Salomon Janhunen}: methodology (equal); writing - review and editing (equal).

\section*{Data availability statement}

The data that support the findings of the study are available from the corresponding author upon reasonable request. 

%%%%%%%%%%%%%%%%%%%%%%%%%%%%%%%%%%%%%%%%%%%%%%%%%%%%%%%%%%
%%%%%%%%%%%%%%%%%%%%%%%%%%%%%%%%%%%%%%%%%%%%%%%%%%%%%%%%%%

\section*{Calcam}\label{sec:calcam}
    
        The temperature information $T_{\text{IR}}$ from the IR camera view (Fig. \ref{fig:bundle-b}) belongs to the 2D pixel space, i.e. $T_{\text{IR}} = T_{\text{IR}}(\phi_{\text{p}}, x_{\text{p}})$, with $\phi_{\text{p}} \in [1;612]$ and $x_{\text{p}} \in [1;540]$ (Sec. \ref{sec:ir_camera}). During the pre-processing phase, Calcam \cite{calcam}, a Python package for geometric calibration of cameras and for performing related analysis, is employed to map such information in the three-dimensional CAD space ($\sim$ real space).
        \\The planar sub-space $\phi x$ of the CAD surface of the divertor tiles (Sec. \ref{sec:dimensionality}) is ultimately of interest in FAHF. Calcam therefore creates the univocal relationship:
        \begin{equation}\label{eq:calcam_map}
            \text{Calcam: $T_{\text{IR}}(\phi_{\text{p}}, x_{\text{p}}) \mapsto $ CAD space $ \mapsto T_{\text{IR}}(\phi, x)$.}
        \end{equation}
        The user input is required when building Eq. (\ref{eq:calcam_map}) via the Calcam GUI, and changes as a function of the camera location and alignment.
        \\Calcam also conveniently allows: (i) to extract useful data related to the camera LoS, FoV and pupil position (Sec. \ref{sec:ir_camera}), among others; (ii) to account for distortion from rectilinear lenses (e.g. barrelling) and fisheye lenses; (iii) to quantify the camera movement.

    \section*{Experimental uncertainty}\label{sec:noise}
    
        This appendix explores the effect of two different flavours of experimental uncertainty of magnitude up to $\eta^n = 0.02 \times \text{max}_{i \in [1;N_x]}\{T_{\text{IR},i}^n\}$ on the experimental datum $T_{\text{IR},i}^{n}$, fed as an input to FAHF. Specifically:
        \begin{itemize}
        
            \item A random noise within $\pm \eta^n$ is added to $T_{\text{IR},i}^n$. The peculiarly smooth behaviour in space of $T_{\text{IR}}$ in Fig. \ref{fig:bundle-c} actually suggests that a random noise would be of even lower entity.
            
            \item A constant shift $+\eta^n$ is applied to $T_{\text{IR},i}^n$, which can be meant as mimicking a calibration/emissivity correction. 
            
        \end{itemize}
        In both cases, an average discrepancy of $2\%$ in the resulting surface heat flux profile in space is found at $t_* = 128$ ms. Crucially, this does not significantly alter its functional shape (Eich function, Sec. \ref{sec:heat_flux}), with the fitting parameters in fact changing by $7\%$ at most $-$ a figure expected to drop further if the iterative fitting procedure of Ref.~\cite{marsden2024} was employed (Sec. \ref{sec:heat_flux_theory}).
        \\Caveated by the above, a nominal variation of $10\%$ in $\lambda_q$ (still acceptable) is found if the magnitude of the constant shift is increased to $8\%$ ($+31$ \textcelsius{}), with an average discrepancy of 5\% over the entire heat flux profile.
        \\In conclusion, a reasonable experimental uncertainty on $T_{\text{IR}}$ would not impact any physical conclusions.

    \section*{Python3 implementation}\label{sec:python}
        
        The entirety of the FAHF algorithms is written in Python3 \cite{python}, also because of its versatility when interfacing with third-party softwares (e.g. Calcam \cite{calcam}, Appendix \ref{sec:calcam}). A modular approach is selected, to accommodate future improvements to the code.
        \\Acknowledging the usual lack of implementation-specific information in the literature dealing with IR thermographical inversions, the actual implementation of Eq. (\ref{eq:heat_eq_fahf_discretised}) in Python3 is here concisely reported.
        \\The meaning of the symbols is obvious, in fact coherent with the mathematical notation adopted throughout the paper. The only exception is for the indeces which are diminshed by 1, because of the Python3 convention of indexing from 0.

\begin{python}

import numpy as np
from scipy.signal import convolve2d

# Fourier numbers for x and y
Fo_x = alpha * dt / dx**2
Fo_y = alpha * dt / dy**2

# 5-point computational stencil
stencil = np.array([
    [ 0,          Fo_y,      0    ],
    [ Fo_x, -2*Fo_x -2*Fo_y, Fo_x ],
    [ 0,          Fo_y,      0    ]
    ])

# initialise temperature array: T = T(x,y,t)
T = np.zeros((Nx, Ny, Nt))

# initial condition @n=0: T_IR = T_IR(x,t)
T[:,:,0] = T_IR[:,0].mean()

for n in range(1,Nt+1):

    # Dirichlet BC @j=0
    T[:,0,n] = T_IR[:,n]

    # explicitly compute temperature rise
    # mode = 'valid' to avoid boundary nodes
    T_rise = convolve2d(in1  = T[:,:,n-1],
                        in2  = stencil,
                        mode = 'valid')
    
    # explicitly compute temperature @n
    # interior nodes only [1:-1,1:-1]
    T[1:-1,1:-1,n] = T[1:-1,1:-1,n-1] + T_rise
        
\end{python}
            
        The code above exemplifies the implementation for interior and Dirichlet nodes. Neumann boundary nodes merely exploits a different shape of the stencil according to Ref.\cite{incropera}, and \texttt{mode = 'same'} in \texttt{convolve2d}. Because of the linearity of the convolution \texttt{convolve2d}, the contribution of the boundary nodes can be separately added to \texttt{T[:,:,n]}.
        \\Notably, \texttt{scipy}'s function \texttt{convolve2d} used to compute derivatives\cite{kernel} is easily parallelised across multiple cores, and even in its serial fashion (30 s per sub-chord, Sec. \ref{sec:space_time_convergence}) allows a 50$\times$ speed-up compared to a for-loop based approach.

    \section*{Details of time and space convergence}\label{sec:convergence_details}

        This appendix further expands on the details of the convergence procedures of which in Sec. \ref{sec:convergence_error}. In the time convergence, the time-step $\Delta t$ is varied in the interval $[4 \times 10^{-6} ; 4 \times 10^{-4}]$ s. The actual values are detailed in Table \ref{tab:time_convergence}. Instead, Table \ref{tab:space_convergence} reports the values $\Delta x \in [6.8 \times 10^{-5} ; 2.3 \times 10^{-3}]$ m of the space convergence. Linear interpolation is employed whenever $\Delta x < \Delta x_{\text{IR}} = 2.1 \times 10^{-4}$ m (Sec. \ref{sec:ir_camera_hardware}), and all throughout the time convergence, as $\Delta t$ from Eq. (\ref{eq:stability}) always falls short of $\Delta t_{\text{IR}} = 1.25 \times 10^{-3}$ s. This does not influence the conclusion of the analyses.
        \\Each simulation in the time convergence is run with a fixed spatial discretisation with $\Delta x= 5.5 \times 10^{-4}$ m. This space-step is small enough such that to not alter the error in time across all the cases, i.e. $o(\Delta x^2) \ll o(\Delta t) \; \forall \Delta t$ in the time convergence. In the space convergence, $\Delta t = 7.2 \times 10^{-6}$ s dually guarantees $o(\Delta t) \ll o(\Delta x^2) \; \forall \Delta x$.
        \\Common to both procedures, a reference point in time $t_* = 128$ ms and a reference point in space $(x_{*},y_{*}) = (16.55 ; 2.21)$ mm are selected where to locally compare the temperature across the various discretisations. The time point $t_* = 128$ ms lies well within $[90;140]\;\text{ms}$, the time window of most interest (Sec. \ref{sec:ref_shot}). Although a convergence analysis is completely agnostic to it, the lack of camera saturation along CH\#2 at this time is instead crucial for the analyses of Secs. \ref{sec:sensitivity_Eich} and \ref{sec:sensitivity_Eich_dy}. Unless otherwise specified, the results obtained at $t = t_*$ apply to any other time instant.
        \\The point $(x_{*},y_{*})$ lies in the volume of the domain, within 3.3 mm of the plasma-facing surface so as to be appreciably affected by a temperature variation (Sec. \ref{sec:theory}), and away from any other boundaries, making it suitable for comparison across different cases. This "local" approach is preferred over integral errors (in the fashion of Sec. \ref{sec:heat_flux_error}) to minimise any near-boundary artifacts which would impair a precise convergence analysis.

    \section*{Algebra of the energy balance error}\label{sec:algebra_energy}

        By neglecting any discretisation error $o(\Delta \xi)$ with $\xi \in \{ x; y; t\}$ in Eqs. (\ref{eq:energy_int}) and (\ref{eq:energy_surf}), it must follow that:
        \begin{equation}\label{eq:int_equal_surface}
            \begin{split}
                \mathcal{E} = & \sum_{i=1}^{N_x} \sum_{j=1}^{N_y} \varrho c_{\text{p}} (T_{i,j}^{n} - T_{i,j}^{n-1}) \times \Delta x \Delta y = \\
                = & \sum_{i=1}^{N_x} q_{i}^{n} \times \Delta x \Delta t \:.
            \end{split}
        \end{equation}
        Additionally, the errors themselves are such that:
        \begin{equation}\label{eq:errors_properties}
            \begin{split}
                & o(\Delta\xi) = \sum_{i} A_i \: o(\Delta\xi) \;\; \forall A_i \in \mathbb{R} \:, \\
                & o(\Delta\xi^0) = o(1) \sim |B| \;\; \forall B \in \mathbb{R} \:, \\
                & o(\Delta\xi^m) \ll o(\Delta\xi^l) \leq o(1) \;\; \forall m > l \in \mathbb{N} \:,
            \end{split}
        \end{equation}
        Therefore, substituting Eqs. (\ref{eq:energy_int}) and (\ref{eq:energy_surf}) in Eq. (\ref{eq:error_energy_time}), and leveraging Eqs. (\ref{eq:int_equal_surface}) and (\ref{eq:errors_properties}) yields:
        \begin{equation}\label{eq:error_scaling}
            \begin{split}
                & \varepsilon_*(\mathcal{E}) = \frac{\left| \Delta\mathcal{E}_{\text{int*}} - \mathcal{E}_{\text{surf*}} \right|}{\left| \Delta\mathcal{E}_{\text{int*}} + \mathcal{E}_{\text{surf*}} \right| / 2} = \\
                & = \frac{|\left[ o(\Delta x^2) + o(\Delta y^2) + o(\Delta t) \right] - \left[ o(\Delta x^2) + o(\Delta y) + o(\Delta t) \right]|}{|\mathcal{E} + \left[ o(\Delta x^2) + o(\Delta y^2) + o(\Delta t) \right] + \left[ o(\Delta x^2) + o(\Delta y) + o(\Delta t) \right]|} \\
                & \sim o(\Delta x^2) + o(\Delta y) + o(\Delta t) \:.
            \end{split}
        \end{equation}
        Finally, the properties specific to time/space convergence of Sec. \ref{sec:convergence_details}, with $\Delta x = \Delta y \Rightarrow o(\Delta x^2) \ll o(\Delta y)$, lead to the relationships:
        \begin{equation}\label{eq:energy_time_convergence_trend}
            \begin{split}
                & \text{time  convergence: $o(\Delta x^2) \ll o(\Delta t)$} \\
                & \Rightarrow \varepsilon_*(\mathcal{E}) \sim o(\Delta y) + o(\Delta t) \:;
            \end{split}
        \end{equation}
        \begin{equation}\label{eq:energy_space_convergence_trend}
            \begin{split}
                & \text{space convergence: $o(\Delta t) \ll o(\Delta x^2)$} \\
                & \Rightarrow \varepsilon_*(\mathcal{E}) \sim o(\Delta y) \:.
            \end{split}
        \end{equation}

%%%%%%%%%%%%%%%%%%%%%%%%%%%%%%%%%%%%%%%%%%%%%%%%%%%%%%%%%%
%%%%%%%%%%%%%%%%%%%%%%%%%%%%%%%%%%%%%%%%%%%%%%%%%%%%%%%%%%

\bibliographystyle{ieeetr}
\bibliography{references}

\end{document}